\documentclass[a4paper,11pt]{article}

\pdfoutput=1
\usepackage{jcappub}

\usepackage{xcolor}	    
\usepackage{hyperref}
\usepackage{listings}
\usepackage{mathtools}
\usepackage{amsmath}
\usepackage{amsfonts}
\usepackage{amssymb}
\usepackage{bbold}
\usepackage{epsfig}
\usepackage{graphicx}
\usepackage{bm}
\usepackage{array}
\usepackage{hyperref}
\usepackage{listings}
\usepackage{color}
\usepackage{float}
\usepackage[normalem]{ulem}
\usepackage{placeins}
\usepackage{bbold}
\usepackage{latexsym}

\newcommand{\GF}{G_{\rm F}}

\newcommand{\jnue}{j_{\nu_e}}
\newcommand{\jnueb}{j_{\bar{\nu}_e}}

\newcommand{\vD}{\boldsymbol{D}}
\newcommand{\vS}{\boldsymbol{S}}
\newcommand{\vect}[1]{\boldsymbol{#1}}

\newcommand{\exclude}[1]{}

\usepackage{tikz,xcolor,hyperref}

\definecolor{lime}{HTML}{A6CE39}
\DeclareRobustCommand{\orcidicon}{\hspace{-1mm}
	\begin{tikzpicture}
	\draw[lime, fill=lime] (0,0) 
	circle [radius=0.16] 
	node[white] {{\fontfamily{qag}\selectfont \tiny \,ID}};
	\draw[white, fill=white] (-0.0525,0.095) 
	circle [radius=0.007];
	\end{tikzpicture}
	\hspace{-3mm}
}

\foreach \x in {A, ..., Z}{\expandafter\xdef\csname orcid\x\endcsname{\noexpand\href{https://orcid.org/\csname orcidauthor\x\endcsname}
			{\noexpand\orcidicon}}
}

\title{Symmetry breaking due to multi-angle matter-neutrino resonance in neutron star merger remnants}

\author[a]{Ian Padilla-Gay\orcidA{},}

\author[b]{Shashank Shalgar\orcidB{},}

\author[b]{and Irene Tamborra\orcidC{}}

\affiliation[a]{SLAC National Accelerator Laboratory, 2575 Sand Hill Rd, Menlo Park, CA 94025}

\affiliation[b]{Niels Bohr International Academy \& DARK, Niels Bohr Institute,\\University of Copenhagen, Blegdamsvej 17, 2100 Copenhagen, Denmark}

\emailAdd{ianpaga@slac.stanford.edu}
\emailAdd{shashank.shalgar@nbi.ku.dk}
\emailAdd{tamborra@nbi.ku.dk}

\abstract{
Neutron star merger remnants are unique sites for exploring neutrino flavor conversion in dense media. Because of the natural excess of $\bar{\nu}_e$ over $\nu_e$,  the neutrino-neutrino potential can cancel the matter potential, giving rise to matter-neutrino resonant flavor conversion. Under the assumption of two (anti)neutrino flavors and spatial homogeneity, we solve the neutrino quantum kinetic equations to investigate the occurrence of the matter-neutrino resonance within a multi-angle framework. We find that isotropy is broken spontaneously, regardless of the mass ordering. Relying on a hydrodynamical simulation of a binary neutron star merger remnant with a black hole of $3\ M_\odot$ and an accretion torus of $0.3\ M_\odot$, we find that complete flavor conversion caused by the matter-neutrino resonance is unlikely, although the matter and neutrino potentials cancel at various locations above the disk. Importantly, the matter-neutrino resonant flavor conversion crucially depends on the shape of the neutrino angular distributions. Our findings suggest that an accurate modeling of the neutrino angular distributions is necessary to understand flavor conversion physics in merger remnants, its implications on the disk physics and synthesis of the elements heavier than iron. 
}

\begin{document}

\hfill SLAC-PUB-17761

\maketitle

\section{Introduction}
\label{sec:intro}

Core collapse supernovae and neutron star mergers are among the richest environments in neutrinos. In neutron star merger remnants, neutrinos are responsible for affecting the disk cooling rate~\cite{Burns:2019byj} as well as the nucleosynthesis of the elements heavier than iron, especially in the polar region~\cite{Janka:2022krt,Metzger:2019zeh}.  Neutrinos change flavor as they propagate, and the rate of flavor transformation can be modified by the medium through which neutrinos travel; a known example of this sort is the Mikheyev-Smirnov-Wolfenstein (MSW) effect in which the neutrino flavor transformation is modified by the coherent forward scattering of neutrinos with the electrons in the medium~\cite{Wolfenstein:1977ue, Mikheev:1986gs}. 
In core-collapse supernovae and neutron star mergers, the neutrino number density is so large that neutrino flavor transformation can also be affected by the coherent forward scattering of neutrinos off other neutrinos.  This causes the flavor of neutrinos and antineutrinos with different momenta to evolve in a correlated manner;
 unlike the MSW effect, this is a nonlinear phenomenon with a rich phenomenology~\cite{Duan:2010bg,Mirizzi:2015eza,Tamborra:2020cul,Richers:2022zug,Chakraborty:2016yeg}.

While neutrino self-interaction can occur in both supernovae and neutron star mergers,  there is a crucial difference between these two environments. In neutron star mergers, there is an excess of electron antineutrinos over neutrinos, while the opposite is true for core-collapse supernovae. The fact that the neutrino self-interaction strength has an opposite sign compared to the neutrino-matter interaction one in neutron star mergers can lead to a cancellation between the two potentials; this phenomenon is called  ``matter-neutrino resonance'' (MNR)~\cite{Malkus:2012ts,Malkus:2014iqa,Malkus:2015mda,Wu:2015fga,Zhu:2016mwa,Vaananen:2015hfa,Frensel:2016fge,Tian:2017xbr,Shalgar:2017pzd,Vlasenko:2018irq}. Due to the MNR, electron neutrinos completely convert into non-electron flavors, while antineutrinos return to their initial state after the MNR transition. This resonant behavior, however, holds in the single-angle (SA) approximation, while only a few investigations on the MNR physics have been carried out within the multi-angle (MA) framework~\cite{Vlasenko:2018irq, Shalgar:2017pzd}. Yet, appropriate modeling of the flavor outcome due to the MNR can be important to predict the nucleosynthetic production~\cite{Janka:2022krt}, as well as contribute to place robust bounds of non-standard physics linked to neutrinos, e.g.~\cite{Sigurdarson:2022mcm}. 

In this paper, we demonstrate that, in an isotropic neutrino gas, the MNR leads to spontaneous breaking of isotropy. Unlike in the case of neutrino self-interaction without MNR~\cite{Cirigliano:2017hmk}, this spontaneous breaking of isotropy occurs for both neutrino mass orderings. We further explore such findings relying on the outputs of a 2D hydrodynamical simulation of a post-merger black hole accretion disk~\cite{Just:2014fka}. 

This work is organized as follows. In Sec.~\ref{sec:eoms}, we introduce the neutrino equations of motion (EOMs). In Sec.~\ref{sec:mnr}, we outline the concept of MNR transitions and highlight the key differences in the MNR occurrence between the SA and MA frameworks. In Sec.~\ref{sec:BHR}, we present the post-merger BH accretion disk system and investigate the occurrence of MNR transitions. Lastly, in Sec.~\ref{sec:conclusions}, we present our conclusions and closing remarks. In addition,  basic physics on the MNR in the SA approximation is covered in Appendix~\ref{sec:single}. Details on the numerical convergence of our simulations are provided in Appendix~\ref{sec:resolution}. Lastly, in Appendix~\ref{sec:mono_dipo}, we reproduce well-known results for the stability conditions of symmetry-breaking solutions in the linear regime.


\section{Neutrino equations of motion}\label{sec:eoms}

In this section, we introduce the neutrino EOMs as well as the MNR physics.

\subsection{Quantum kinetic equations}
 For the sake of simplicity, we rely on the two-flavor approximation:  $(\nu_e, \nu_x)$. The neutrino field can then be  modeled in terms of $2\times2$ density matrices, $\rho(\vec{x},\vec{p},t)$ for neutrinos and $ \bar{\rho}(\vec{x},\vec{p},t)$ for antineutrinos. 
The neutrino and antineutrino density matrices are respectively,
\begin{eqnarray}
\rho(\vec{x}, \vec{p},t) = 
\begin{pmatrix}
\rho_{ee}(\vec{x}, \vec{p},t) & \rho_{ex}(\vec{x}, \vec{p},t)\\
\rho_{ex}^{*}(\vec{x}, \vec{p},t) & \rho_{xx}(\vec{x}, \vec{p},t) 
\end{pmatrix}\ \ \mathrm{and}\ \ 
\bar{\rho}(\vec{x}, \vec{p},t) = 
\begin{pmatrix}
\bar{\rho}_{ee}(\vec{x}, \vec{p},t) & \bar{\rho}_{ex}(\vec{x}, \vec{p},t)\\
\bar{\rho}_{ex}^{*}(\vec{x}, \vec{p},t) & \bar{\rho}_{xx}(\vec{x}, \vec{p},t) 
\end{pmatrix}\ ,
\end{eqnarray} 
where the diagonal elements represent the occupation numbers of neutrinos of different flavors, while the off-diagonal terms contain information on flavor coherence. Since the luminosity of the non-electron flavors is negligible with respect to the ones of $\nu_e$ and $\bar\nu_e$ in the BH torus disk~\cite{Just:2015dba,Janka:1999qu,Sekiguchi:2016bjd}, we consider  $\rho_{xx}(t=0)=\bar{\rho}_{xx}(t=0)=0$ hereafter. 

The  EOMs for neutrinos are
\begin{eqnarray}
\label{eq:eoms}
\kern-1em (\partial_t+\vec{v}\cdot \vec{\nabla}) \rho(\vec{x},\vec{p},t) &=&-i [H(\vec{x},\vec{p},t),\rho(\vec{x},\vec{p},t)] 
+ \mathcal{C}[\rho(\vec{x}, \vec{p},t)] \ ;
\end{eqnarray}
 the term on the left-hand side is the total derivative which includes the advective term ($\vec{v}\cdot \vec{\nabla}$)  that is neglected throughout this paper---see Refs.~\cite{Shalgar:2019qwg, Padilla-Gay:2020uxa} for the role of advection in the context of flavor conversion.  The last term ($\mathcal{C}$) on the right-hand side is the collision term, including emission, absorption, and momentum-changing scatterings of neutrinos with matter~\cite{Shalgar:2022rjj, Shalgar:2022lvv, Padilla-Gay:2022wck, Fiorillo:2023ajs}; this term is also neglected in the rest of this work for simplicity. Note, however, that the advective and collision terms determine the angular distribution of neutrinos; flavor evolution further contributes to modifying the angular distributions. Throughout this work, we use natural units ($\hbar=c=1$).

The Hamiltonian, $H(\vec{x},\vec{p},t)$, that governs the EOMs  has three  terms:  the vacuum, matter, and the self-interaction terms:
\begin{eqnarray}
\label{Hamiltonian}
H(\vec{x},\vec{p},t) = H_{\mathrm{vac}}(\vec{p}) + H_{\mathrm{mat}}(\vec{x},\vec{p}) + H_{\nu\nu}(\vec{x},\vec{p},t)\ .
\end{eqnarray}
A similar equation of motion holds for antineutrinos, except for  $H_{\mathrm{vac}} \rightarrow -H_{\mathrm{vac}}$. 
The vacuum term is 
\begin{eqnarray}
    H_{\mathrm{vac}} = 
    \frac{\omega}{2} \begin{pmatrix}
        -\cos{2\theta_V} & \sin{2\theta_V} \\
        \sin{2\theta_V} & \cos{2\theta_V}
    \end{pmatrix} \ , 
\end{eqnarray}
where $\omega = \Delta m^2/2E$ is the vacuum oscillation frequency and $\theta_V$ is the vacuum mixing angle. We assume that all neutrinos have the same energy for the sake of simplicity and set  $\omega \equiv {\Delta m^{2}}/{2E} = 1\ \textrm{km}^{-1}$. Note that, in order to trigger the MNR, sufficiently large values of the mixing angle are required (see Appendix~\ref{sec:single}). Hence, we use $\theta_{\textrm{V}} = 0.15\ \textrm{rad}$, unless otherwise specified. The matter Hamiltonian due to the coherent forward scattering with the background matter is given by
\begin{eqnarray}
    H_{\mathrm{mat}} = \lambda \begin{pmatrix}
        1 & 0 \\
        0 & 0 \\
    \end{pmatrix} \ , 
\end{eqnarray}
where the matter potential, 
\begin{eqnarray}\label{eq:lambda}
    \lambda = \frac{\sqrt{2}\GF \rho_B}{m_N} Y_e \ ,
\end{eqnarray}
depends on the Fermi constant $\GF$, the baryon mass density $\rho_B$, the nucleon mass $m_N$, and the electron fraction $Y_e=(n_{e^{-}}-n_{e^{+}})/n_B$. 
The neutrino-neutrino interaction term of the Hamiltonian takes into account  the coherent forward scattering of neutrinos with their own background and is given by
\begin{eqnarray}\label{eq:hnunu}
    H_{\nu\nu}(\vec{p},t) = \mu \int d\vec{v}^\prime[\rho(v^\prime,t)-\bar{\rho}(v^\prime,t)][1-\vec{v}\cdot \vec{v}^\prime]\ , 
\end{eqnarray}
where $\mu$ is the neutrino-neutrino interaction strength. The $H_{\nu\nu}$ term couples neutrinos of different momenta and is responsible for the non-linear nature of neutrino flavor conversion. The case with $\vec{v}\cdot \vec{v}^\prime=0$ corresponds to the SA approximation (Eq.~\ref{eq:hnunu} reduces to $H^{\mathrm{SA}}_{\nu\nu}(t)=2\mu[\rho(t)-\bar{\rho}(t)]$); notice that one obtains the same reduced system from Eq.~\ref{eq:hnunu} choosing  isotropic $\rho(v,t)$ and $\bar{\rho}(v,t)$ so that the integral of the term proportional to $\vec{v}\cdot \vec{v}^\prime$ vanishes~\cite{Fogli:2007bk,Duan:2006an}. The term $(1-\vec{v}\cdot\vec{v}^{\prime})$ in Eq.~\ref{eq:hnunu} depends on both polar and azimuthal angles, $\theta$ and $\phi$, respectively. However, for the sake of simplicity, we ignore the dependence on $\phi$, assuming axial symmetry. The consequences of this simplification are not known for the case of MNR, however, they might be non-trivial as found in the context of fast flavor evolution~\cite{Shalgar:2021oko}.

\subsection{Matter-neutrino resonance}
The MNR  is the enhancement of flavor conversion due to the cancellation in the diagonal terms of the Hamiltonian:
\begin{eqnarray}
\label{rescond}
H_{ee}^{\nu\nu}(\vec{x},\vec{p})-H_{xx}^{\nu\nu}(\vec{x},\vec{p})+\lambda(\vec{x}) \approx 0 \ .
\end{eqnarray}
Since the self-interaction Hamiltonian ($H_{\nu\nu}$) is angle-dependent, this condition cannot be satisfied for all angles simultaneously in the MA case, unless the angular distributions are isotropic. When the SA approximation is relaxed, the MNR  can occur at any points in space and momentum modes for which Eq.~\ref{rescond} is satisfied. As time evolves, the advection of neutrinos (both forward and backward according to the direction of propagation) ensures that the flavor-evolved neutrinos affect the flavor composition at other spatial locations. 

The full solution of the EOMs, in the presence of the MNR and as a boundary problem involving collisions and advection, is beyond the reach of present-day computational facilities.
In order to gain an insight into the difference between neutrino flavor evolution due to the MNR in the SA and MA cases, in Sec.~\ref{sec:mnr}, we rely on a simplified system with angular distributions fixed a priori and with the strength of the self-interaction potential and the matter potential changing as functions of time, while ignoring the spatial dependence of the latter.



\section{Matter-neutrino resonance in an ideal neutrino gas} \label{sec:mnr}
Before investigating the MNR in the context of neutron star merger remnants, we focus on an idealized neutrino gas. Hereafter, we assume that the spatial coordinate is directly connected to the time coordinate through the speed of light $c$ since neutrino advection and non-forward collisions are neglected. We consider a time-dependent self-interaction potential and constant matter potential: 
\begin{eqnarray}\label{eq:mulam}
\mu(t) = 3\times 10^{3} e^{-qt}\ \textrm{km}^{-1}  \ \ \mathrm{and}\ \ 
\lambda = 10^{3}\ \textrm{km}^{-1}\ ,
\end{eqnarray}
where $q = 0.1$~km$^{-1}$ (or equivalently $3 \times 10^{4}$~s$^{-1}$). We stress that $\mu$ is proportional to the neutrino number density; however, in this section, we choose to define the self-interaction strength as an arbitrary function of time (this simplification is relaxed in Sec.~\ref{sec:BHR}).

For the MNR to occur, Eq.~\ref{rescond} should be satisfied, which is only possible if there are more antineutrinos than neutrinos. Hence, we assume that   $\rho_{ee}(t_0)=1$ and $\bar{\rho}_{ee}(t_0)=4/3$ at the initial time $t_0$, and  the  initial asymmetry parameter is
\begin{eqnarray}\label{eq:alpha}
\alpha \equiv \frac{\bar{\rho}_{ee}(t_0)}{\rho_{ee}(t_0)} = \frac{4}{3}\ .
\end{eqnarray}
With these initial conditions, the neutrino-neutrino Hamiltonian in the SA approximation is $H^{\mathrm{SA}}_{\nu\nu}(t_0)=2\mu[1-\alpha]$. The same  Hamiltonian can be obtained for the MA system with initially isotropic angular distributions. Note that, in the MA case, we solve the EOMs fixing the angular distributions at $t_0$, the neutrino distributions then evolve because of flavor conversion (but not due to advection and collisions, both neglected in this work).

\subsection{Single-angle and multi-angle solutions of the  matter-neutrino resonance for an isotropic neutrino gas}\label{sec:sa}

\begin{figure*}[t!]
\includegraphics[width=0.99\textwidth]{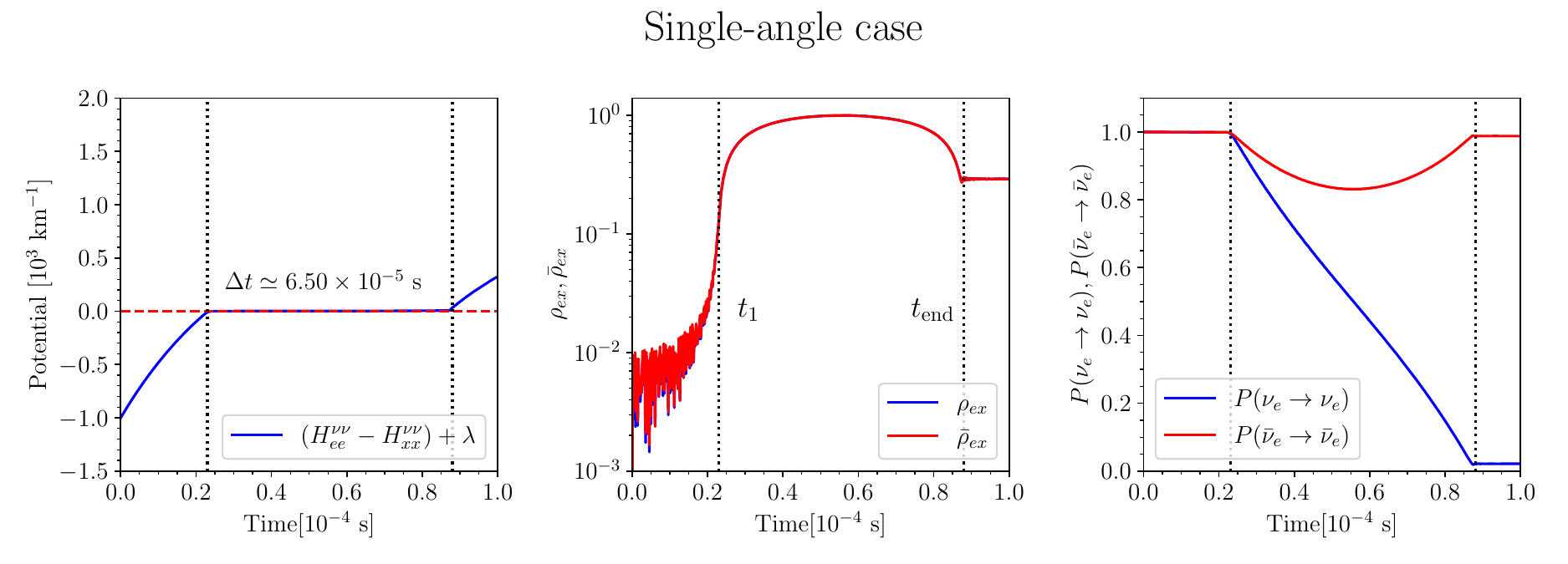}
\includegraphics[width=0.99\textwidth]{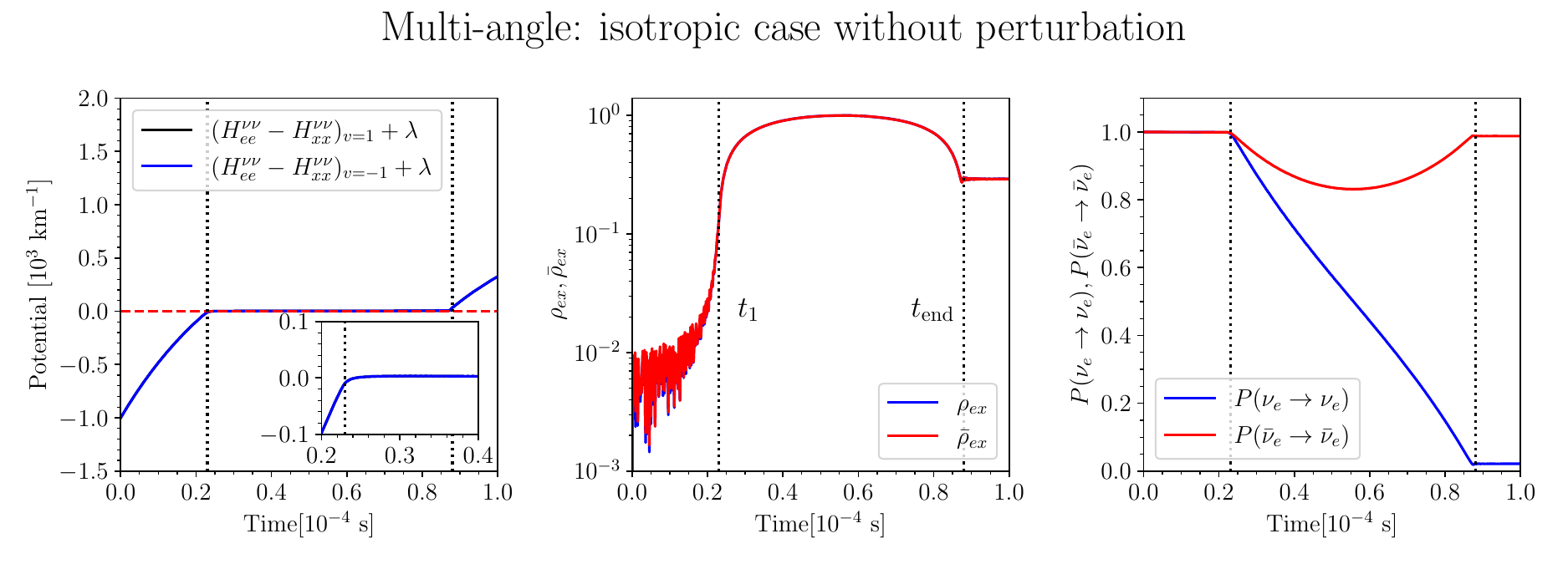}
\includegraphics[width=0.99\textwidth]{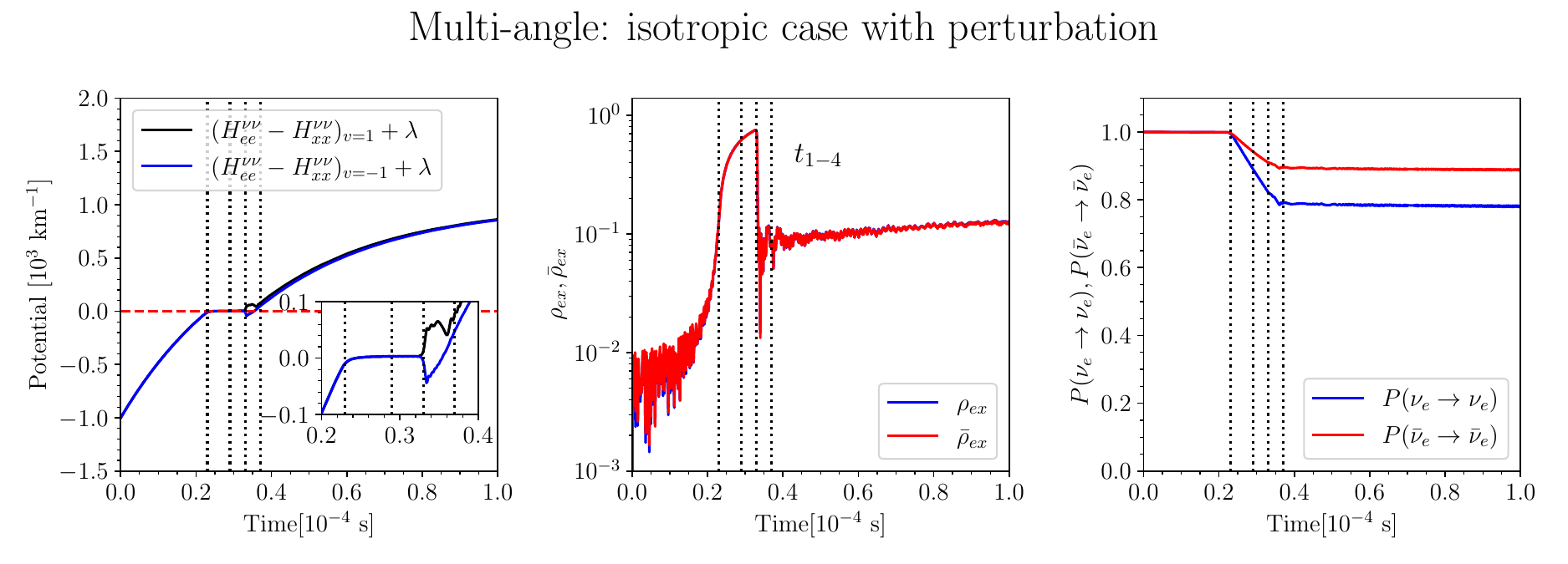}
\caption{Flavor evolution for an ideal neutrino gas. Single angle (top), multi-angle with isotropic angular distributions (middle), and multi-angle with isotropic angular distributions and a perturbation of $10^{-22}\cos{\theta}$  in $\bar{\rho}_{ee}$ (bottom). \textit{Left:} Oscillated total potential (sum of self-interaction and matter potentials) as the MNR transition occurs. \textit{Middle:} Off-diagonal terms of the density matrices,  $\rho_{ex}$ and $\bar\rho_{ex}$.  \textit{Right:}  Survival probabilities for electron neutrinos and antineutrinos.  For the SA case (top panels), the duration of the transition $\Delta t$ can be computed analytically (see Eq.~\ref{eq:deltaT1}). The MA isotropic case without perturbation reproduces the SA approximation; $500$ bits per number are needed to achieve this agreement. For the MA isotropic case with perturbation (bottom panels), the duration of the MNR transition, $\Delta t = t_3-t_1$, is smaller than in the SA approximation. Due to the growth of non-isotropic modes, the MNR transition is interrupted in the MA case around $t_3=0.33\times 10^{-4}$~s and flavor conversion is not complete for electron neutrinos.}
\label{fig:6a}
\end{figure*}

Most of the MNR investigations in the literature have been performed using the SA approximation~\cite{Malkus:2014iqa,Vaananen:2015hfa,Wu:2015fga,Malkus:2015mda,Zhu:2016mwa}. We briefly review the MNR results in the SA approximation in order to compare them with the ones obtained in the MA approach. 
The top panels of Fig.~\ref{fig:6a} show the oscillated total potential $H_{ee}^{\nu\nu}-H_{xx}^{\nu\nu}+\lambda$, the evolution of the off-diagonal terms of the density matrices, and the survival probabilities of $\nu_e$ and $\bar\nu_e$, from left to right respectively. As we can see,  a general feature of the MNR in the SA approximation is that $\nu_e$'s completely change their flavor while $\bar\nu_e$'s return to their original configuration. Unlike the MSW effect, the MNR is independent of the sign of $\omega$; however,  $\theta_V$ should be large enough for the MNR to be possible, as discussed in Appendix~\ref{sec:single} (see also Fig.~\ref{fig:4a}). The times $t_1$ and $t_{\mathrm{end}}$ (see dotted vertical lines in the top panels of Fig.~\ref{fig:6a}) mark the times at which the cancellation of the matter and neutrino-neutrino potentials takes place. During $\Delta t = t_{\mathrm{end}}-t_1 \simeq 6.5 \times 10^{-5}$~s,  $\rho(t)$ and $\bar{\rho}(t)$ evolve holding the MNR conditions. The analytical dependence of  $\Delta t = t_{\mathrm{end}}-t_1$ on the properties of the neutrino gas is provided in Ref.~\cite{Malkus:2014iqa}. A direct comparison between the analytical estimates and our numerical results is shown in Appendix~\ref{sec:single} (cf.~Fig.~\ref{fig:1a}), where one can see that the agreement between both approaches is excellent.

The simplest comparison between the  MA and  SA flavor outcomes can be carried out considering a gas of neutrinos that is initially isotropic. 
As in the SA approximation, the MA Hamiltonian does not have any dependence on $\vec{v}$ initially and naively one would not expect a major difference in the flavor outcome with respect to the SA case. 
 However, there is a stark difference in the flavor evolution between the SA and MA cases.  This can be seen by comparing the three cases shown in the three rows of Fig.~\ref{fig:6a}. For each row, the panels show the evolution of the oscillated potential, the evolution of the off-diagonal components of the density matrices, and the survival probabilities for neutrinos and antineutrinos from left to right. The plots in the top row refer to the SA case. The same angle-averaged quantities are shown for an isotropic system in the middle row. The evolution of all the quantities is identical in the first and the middle rows as expected. However, we see a drastic difference in the evolution in the bottom row compared to the middle row even though the only difference between the two systems is that for the system in the bottom row a perturbation of $10^{-22}\cos\theta$ is added to the antineutrino angular distribution. We note that the size of the perturbation is smaller than the error of a double precision number on computers. Indeed, we use $500$ bits of precision per number and an absolute and relative tolerance of $10^{-25}$ each. If double precision floating point numbers are used, then the isotropy of the system is broken by the floating point error and it is not possible to obtain the results presented in the middle row of Fig.~\ref{fig:6a}~\footnote{We rely on the support for arbitrary precision of numbers in the \texttt{DifferentialEquations.jl} package of \texttt{Julia}~\cite{DifferentialEquations.jl-2017, Julia-2017}.
In the absence of these precautions, the results depend on the solver used to solve the differential equations.}. Hence, the MA case undergoes a spontaneous breaking of isotropy even if a tiny anisotropy is initially present.

The spontaneous breaking of isotropy has been known to exist in collective neutrino flavor conversion~\cite{Duan:2013kba, Cirigliano:2017hmk} for either mass ordering depending on whether there is an excess of neutrinos over antineutrinos or vice versa. However, in the MNR case,  the spontaneous breaking of isotropy is present irrespective of the mass ordering because the underlying MNR transition that triggers the symmetry breaking is almost identical in both mass orderings, as shown in Appendix~\ref{sec:single} (Fig.~\ref{fig:5a}).  Moreover, this spontaneous breaking of isotropy manifests itself only if the isotropic mode is unstable.

If the isotropy is spontaneously broken during the MNR, this is reflected in the time evolution of the survival probability of neutrinos, 
\begin{eqnarray}\label{eq:Psurvival}
    P(\nu_e\rightarrow \nu_e) = \frac{\int dv [\rho_{ee}(v,t) - \rho_{xx}(v,t_0)]}{\int dv [\rho_{ee}(v,t_0)-\rho_{xx}(v,t_0)]} \ 
\end{eqnarray}
(a similar definition holds  for $P(\bar{\nu}_e\rightarrow \bar{\nu}_e)$ with the replacement $\rho_{ii}\rightarrow \bar{\rho}_{ii}$). In the bottom right panel of Fig.~\ref{fig:6a}, one can see that the survival probabilities are affected by the anisotropic modes developing between $t_1$ and $t_3$. Note that the evolution of $P(\nu_e\rightarrow \nu_e)$ for the SA and MA cases are substantially different; this is especially true for $P(\bar{\nu}_e\rightarrow \bar{\nu}_e)$, which does not describe full flavor conversion in the MA case (cf.~right bottom panel of Fig.~\ref{fig:6a}).

\begin{figure}[t!]
\begin{center}
\includegraphics[width=0.99\textwidth]{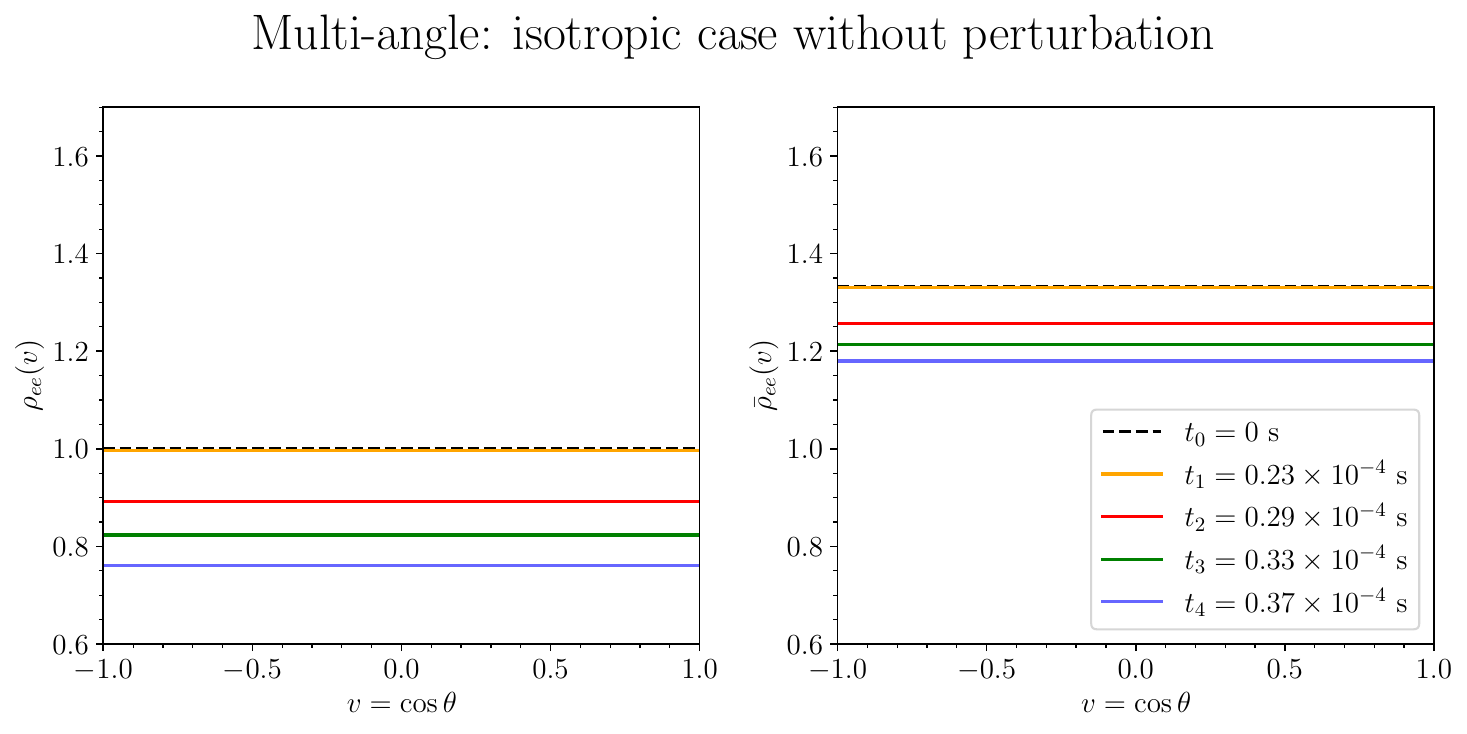}
\includegraphics[width=0.99\textwidth]{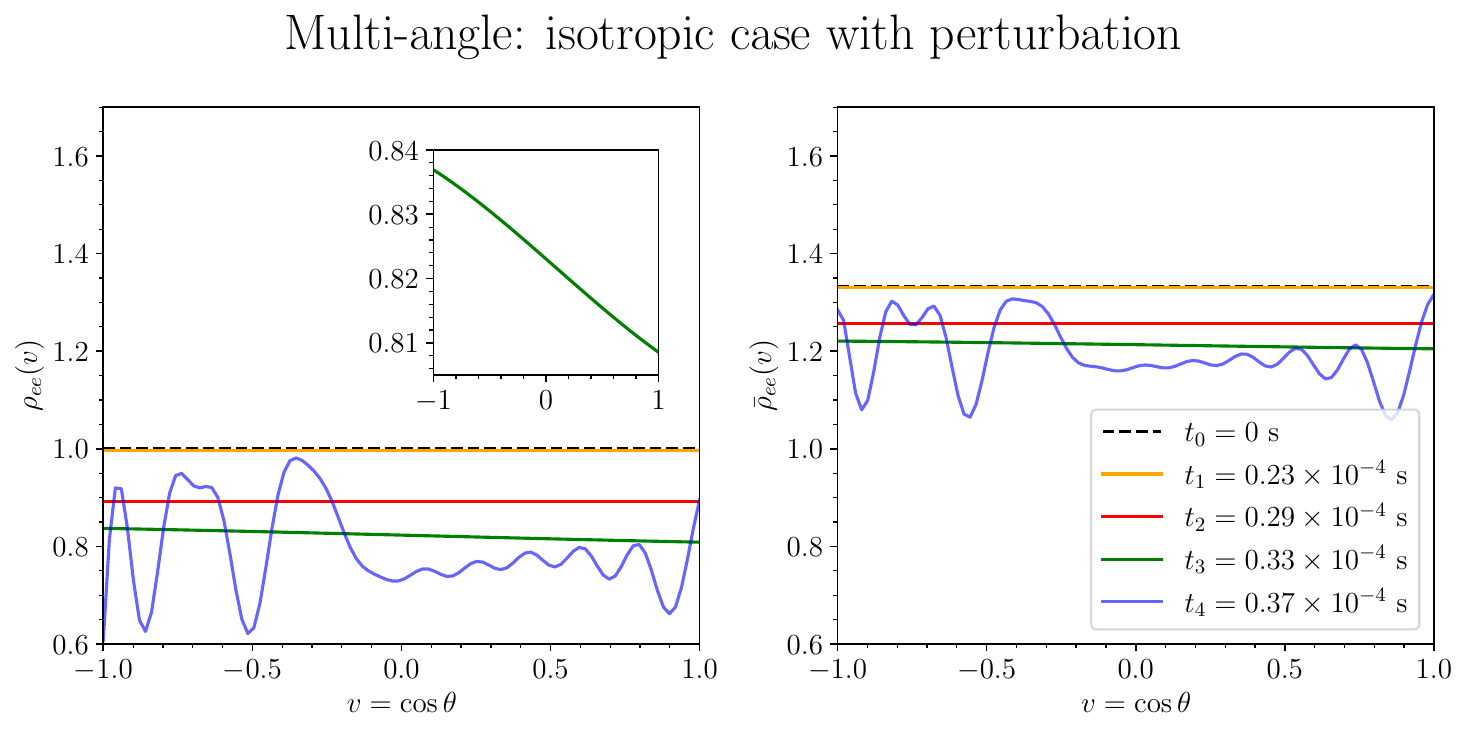}
\end{center}
\caption{Snapshots of the angular distributions of $\rho_{ee}$ (left panels) and $\bar\rho_{ee}$ (right panels) for the MA isotropic case without perturbation (top panels) and MA isotropic case with perturbation (bottom panels) shown in Fig.~\ref{fig:6a}. For the MA case without perturbation (top), the system preserves the isotropy in the neutrino angular distributions during the MNR. For the MA case with perturbation (bottom), the angular distributions start to evolve as the  MNR condition is met. At  $t_3$, a dipole (green curve) forms dynamically breaking isotropy as shown in the inset; at the same time, a ``jump'' in $\rho_{ex}$ occurs as visible from the bottom left panel of Fig.~\ref{fig:6a}. From $t_1$ to $t_4$, diffusion of flavor waves from large to small angular scales takes place.
\label{fig:8a}}
\end{figure}

The sudden jump in the evolution of the off-diagonal terms of the density matrix in the middle bottom panel of Fig.~\ref{fig:6a} can be better understood by looking at the time snapshots of the perturbed $\nu_e$ and $\bar\nu_e$ angular distributions in the bottom panel of Fig.~\ref{fig:8a}, before and after the MNR. The system enters the MNR at about $t_1$ (see Fig.~\ref{fig:6a}), but the MNR is interrupted at $t_3$. The neutrino and antineutrino angular distributions no longer preserve the isotropy that was initially imposed on them. A clear particle excess (deficit) in the backward (forward) direction dynamically forms during the MNR in the MA solution. More precisely, while the MNR condition holds, the $v$-dependence of the density matrices in Fig.~\ref{fig:8a} slowly transitions from isotropic (orange and red curves) to linear in $v$ (green curve), and later to more complex functions non-linear in $v$ (blue curve). The $t_{3}$ and $t_{4}$ snapshots show the angular distributions in the non-linear regime--small-scale angular structures develop within a few $10^{-6}$~s. This diffusion from large to small angular scales is triggered by the MNR. Animations of the temporal evolution of the potentials, survival probabilities and (anti)neutrino angular distributions are provided as \href{https://sid.erda.dk/share_redirect/e2zTyjhG3B/index.html}{Supplemental Material}. In Appendix~\ref{sec:resolution} we show that the breaking of isotropy, and therefore the cascade of flavor waves to small scales, still occurs if the number of angular bins is increased, confirming that this is a MA effect caused by tiny anisotropies of $\mathcal{O}(10^{-22}\cos{\theta})$ and not an artifact caused by insufficient angular resolution.

\begin{figure*}[t!]
\begin{center}
\includegraphics[width=0.90\textwidth]{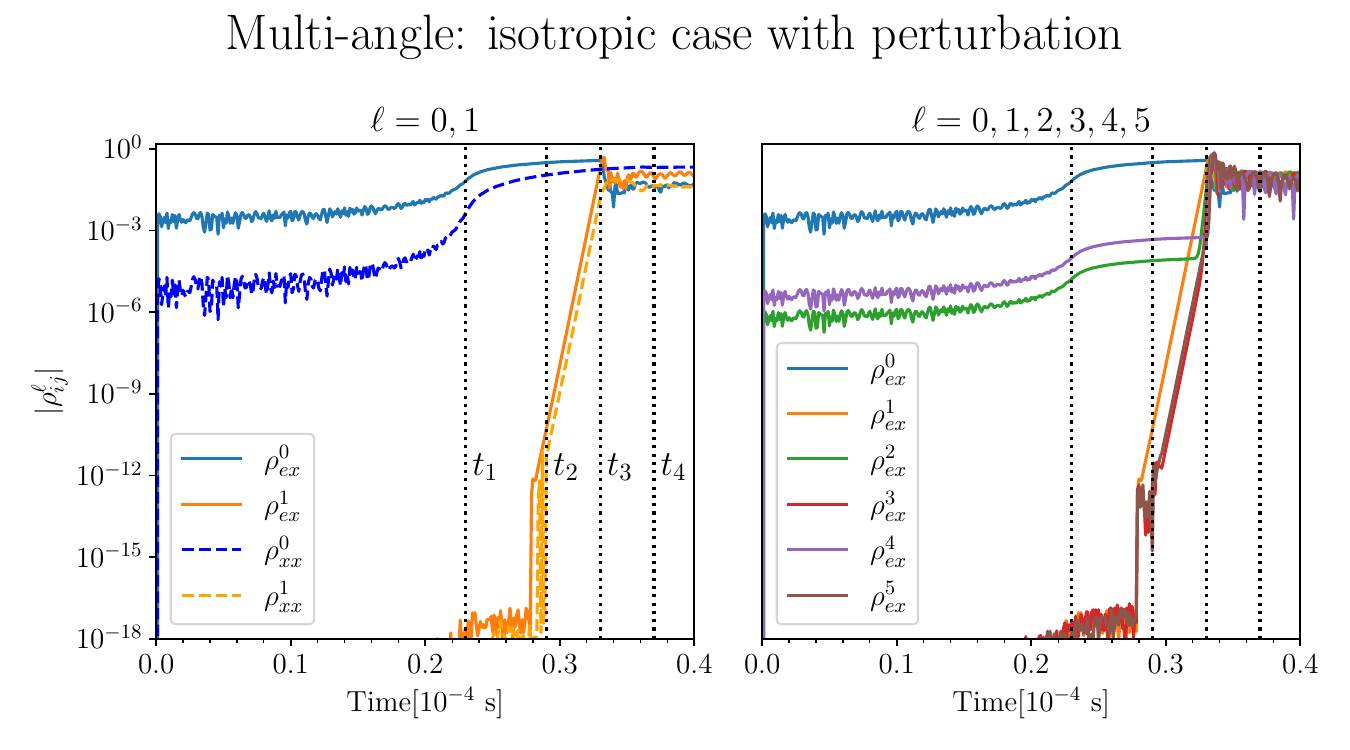}
\end{center}
\caption{Temporal evolution of the Legendre coefficients of  $\rho_{ij}$ for the system in the bottom panels of Fig.~\ref{fig:6a}. \textit{Left: } Time evolution of the norm of $\rho_{ex}^{0}$, $\rho_{xx}^{0}$, $\rho_{ex}^{1}$ and $\rho_{xx}^{1}$. During  $\Delta t=t_3-t_1$, the potentials cancel each other (see bottom panel in Fig.~\ref{fig:6a}), and the system transitions from isotropic to non-isotropic (see bottom panel in Fig.~\ref{fig:8a}) as a result of the growing dipole. Interestingly, during  $\Delta t$, both $\rho_{ex}^1$ and $\rho_{xx}^1$ grow exponentially at the same rate. \textit{Right: } Time evolution of the first six $\ell$ modes. Modes with $\ell>1$ also grow within the time interval $ \Delta t=t_3-t_1$ during the MNR. Larger $\ell$ modes are fully developed around $t_3$, when the evolution is highly non-linear. The discontinuity around $t_2$ is caused by the post-process ($64$ bits) of the numerical solution ($500$ bits) which has a larger floating-point error.
}
\label{fig:9a}
\end{figure*}

In order to understand the breaking of isotropy, it is useful to decompose the (anti)neutrino angular distributions in terms of Legendre polynomials:
\begin{eqnarray}
    \rho_{ij} = \sum_{\ell=0}^{\infty} \rho_{ij}^{\ell} P_{\ell} \ , 
\end{eqnarray}
where $\rho_{ij}^{\ell}$ are the coefficients of the expansion that can be determined by integrating the product of the density matrix elements and the corresponding Legendre polynomial, i.e.
\begin{eqnarray}
    \rho_{ij}^{\ell} = \frac{2\ell+1}{2} \int_{-1}^{1} dv \rho_{ij}(v) P_{\ell}(v) \ .
\end{eqnarray}
Thus $\rho_{ij}(v,t)$ allows us to compute the coefficients $\rho_{ij}^{\ell=0,...,5}(t)$. These coefficients encode information about the structure of the angular distributions and help to gauge the level of anisotropy of the neutrino gas. In the following, we refer to the  $\ell=0$ and $\ell=1$ modes as the ``monopole'' and the ``dipole'' of the angular distributions, respectively.

Figure~\ref{fig:9a} shows the time evolution of the monopole ($\ell=0$) and dipole ($\ell=1$) of  $\rho_{ex}$  in the left panel, as well as the evolution of higher $\ell>1$ modes in the right panel. We can see that  $\rho_{ex}^1$ and $\rho_{xx}^1$ (similar behavior for antineutrinos, although not shown) start at $\sim10^{-18}$ and grow exponentially during  $\Delta t = t_{3}-t_{1}$ (note that it is numerically impossible to have $\rho_{ex}^1$  exactly equal to zero due to the floating-point error---see also Appendix~\ref{sec:resolution}). The discontinuity in Fig.~\ref{fig:9a} where $|\rho_{ex}^{1}|$ jumps from $10^{-18}$ to $10^{-12}$ is because, in the post-processing of the solution of the EOMs, we implement double precision ($64$ bits per number) to compute the Legendre coefficients, while the numerical solution is obtained with $500$ bits per number; the exponential growth would also be visible from $t_1$ and onwards, if the post-processing had the same level of floating-point error accuracy of the numerical EOM solution. During $\Delta t$,  a partial MNR transition occurs (cf.~the time interval between $t_{1}$ and $t_3$ in Fig.~\ref{fig:6a}). As the dipole modes grow during $\Delta t$, the angular distributions become non-isotropic, as shown in Fig.~\ref{fig:8a} at $t_1=0.23\times10^{-4}$~s and $t_3=0.33\times10^{-4}$~s. 
Furthermore, larger $\ell>1$ modes also grow exponentially early on around $t_1$, as displayed in the right panel of Fig.~\ref{fig:9a}. Large $\ell$ modes are not visible in the angular distributions until later times ($t_{3-4}$) because they are initially negligible. In the non-linear regime, not only the $\ell=1$ modes have fully developed but also higher ($\ell>1$) modes. This is a crucial difference with respect to the SA approximation: the coherence between different angular modes is not maintained in the MA case due to the growth of high $\ell$ multipoles during the MNR.

\subsection{Multi-angle solution of the matter-neutrino resonance for a non-isotropic neutrino gas}\label{sec:ma-noniso}

The  MNR cancellation found in the MA isotropic case in Sec.~\ref{sec:sa} is possible due to all angle bins initially acting as one since they share the same $H_{\nu\nu}(v)$ as the MNR begins. We now investigate whether (partial) MNR transitions are likely to occur in a non-isotropic neutrino gas. 
To this purpose, we consider initial angular distributions that  are non-isotropic and parametrize them as follows 
\begin{eqnarray}\label{eq:distr_noniso}
    \rho_{ee}(v) = 1 + a v \ \ \mathrm{and}\ \ 
    \bar{\rho}_{ee}(v) = \alpha (1 + \bar{a} v) \ ,
\end{eqnarray}
with $\alpha=\int dv \bar{\rho}_{ee}/\int dv \rho_{ee}=4/3$ (as for the SA and MA cases in the previous section, see Eq.~\ref{eq:alpha}). The parameters $a$ and $\bar{a}$  provide the initial degree of anisotropy and are arbitrarily fixed to $a=0.1$ and $\bar{a}=0.2$ (note that these distributions do not have electron lepton number crossings~\cite{Tamborra:2020cul}). We define $\mu(t)$, $\lambda$, $\omega$, and $\theta_V$ as in Eq.~\ref{eq:mulam} to facilitate a comparison with our previous SA and MA results.

A crucial difference between the SA and isotropic MA calculations is that the Hamiltonian depends on $(1-v v^\prime)$ in the latter case and couples different angular modes with each other even before the MNR cancellation occurs. As a result, different angle modes have different $H_{ee}^{\nu\nu}(v)-H_{xx}^{\nu\nu}(v)$, which result in different locations for the MNR  as a function of $v$; all possible  $H_{ee}^{\nu\nu}(v)-H_{xx}^{\nu\nu}(v) + \lambda$ for any $v$ lie in between the corresponding values for $v=-1$ and $v=1$.

\begin{figure*}[t!]
\includegraphics[width=0.99\textwidth]{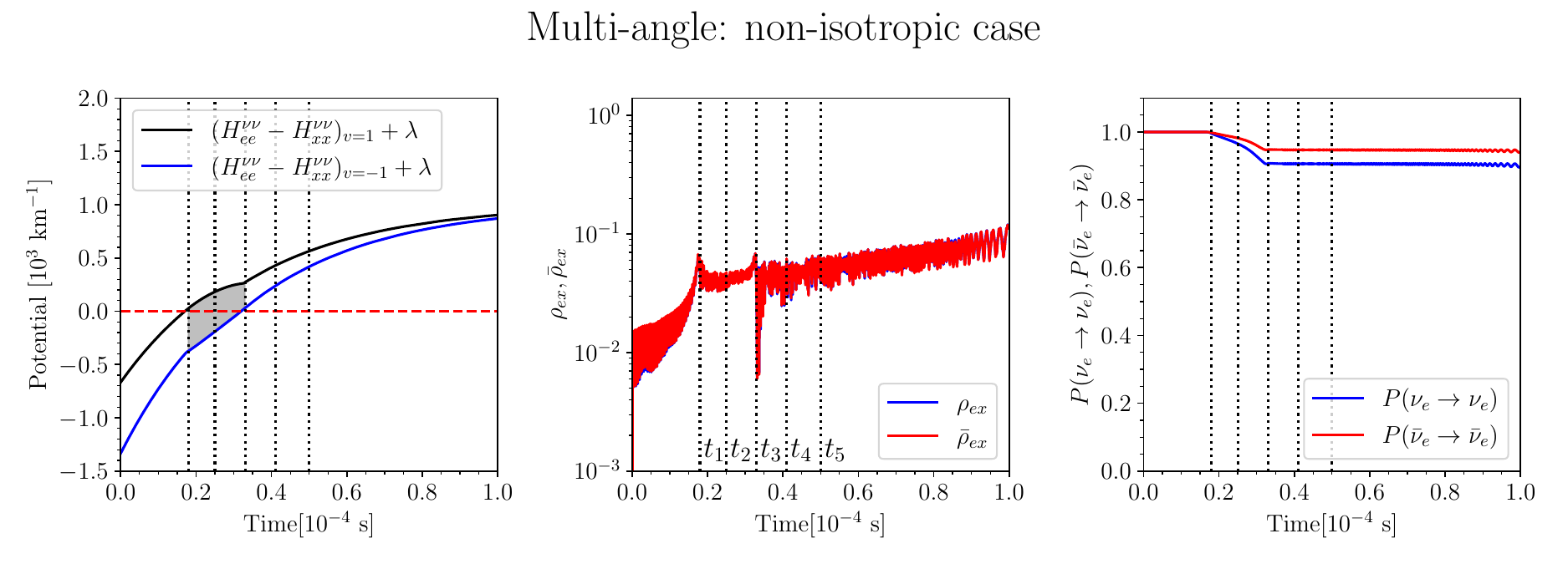}
\caption{Multi-angle solution for an ideal neutrino gas with non-isotropic angular distributions, cf.~Fig.~\ref{fig:6a} for comparison. \textit{Left:} Oscillated total potentials as functions of time. We mark a few selected times, $t_{1-5}$, to highlight when the MNR condition is met for different angular modes. The gray-shaded area marks the time interval during which the MNR occurs. \textit{Middle:} Off-diagonal terms of the density matrices of neutrinos and antineutrinos. \textit{Right:}  Survival probabilities for $\nu_e$'s and $\bar\nu_e$'s. Although the total potential cancels out for all angular modes at different times, none of them successfully undergoes full MNR transition. The oscillated potentials at $v=\pm 1$ (black and blue) do not remain zero (dashed line), which is a characteristic feature of MNR transitions. Snapshots of the (anti)neutrino angular distributions at $t_{1-5}$ are shown in Fig.~\ref{fig:11a}. 
}
\label{fig:10a}
\end{figure*}
Figure~\ref{fig:10a} shows the oscillated potentials for the angular bins $v=\pm 1$ (left panel; the envelope of all potentials obtainable for any $v$ is enclosed between the black and blue lines), the evolution of the off-diagonal entries of the density matrices (middle panel), and the survival probabilities for $\nu_e$ and $\bar\nu_e$ (right panel). The total potential for the $v=1$ bin (black line) fulfills the MNR condition at $t_1$. At $t_3$, the angle bin with $v=-1$ (blue line) fulfills the MNR condition. To guide the eye, the gray-shaded area highlights the region within which any $v$ between  $v=-1$ and $v=1$ meets the MNR condition. Most of the flavor dynamics takes place within the gray-shaded area (which highlights the time interval between $t_1$ and $t_3$); this is clearly visible from the evolution of the survival probabilities in the right panel of Fig.~\ref{fig:10a}. For $t>t_3$, the flavor dynamics seems to come to a halt, and an approximate steady state is reached. These results are qualitatively similar to those reported in Figs.~4 and 5 of Ref.~\cite{Vlasenko:2018irq}.

\begin{figure}[t!]
\begin{center}
\includegraphics[width=0.99\textwidth]{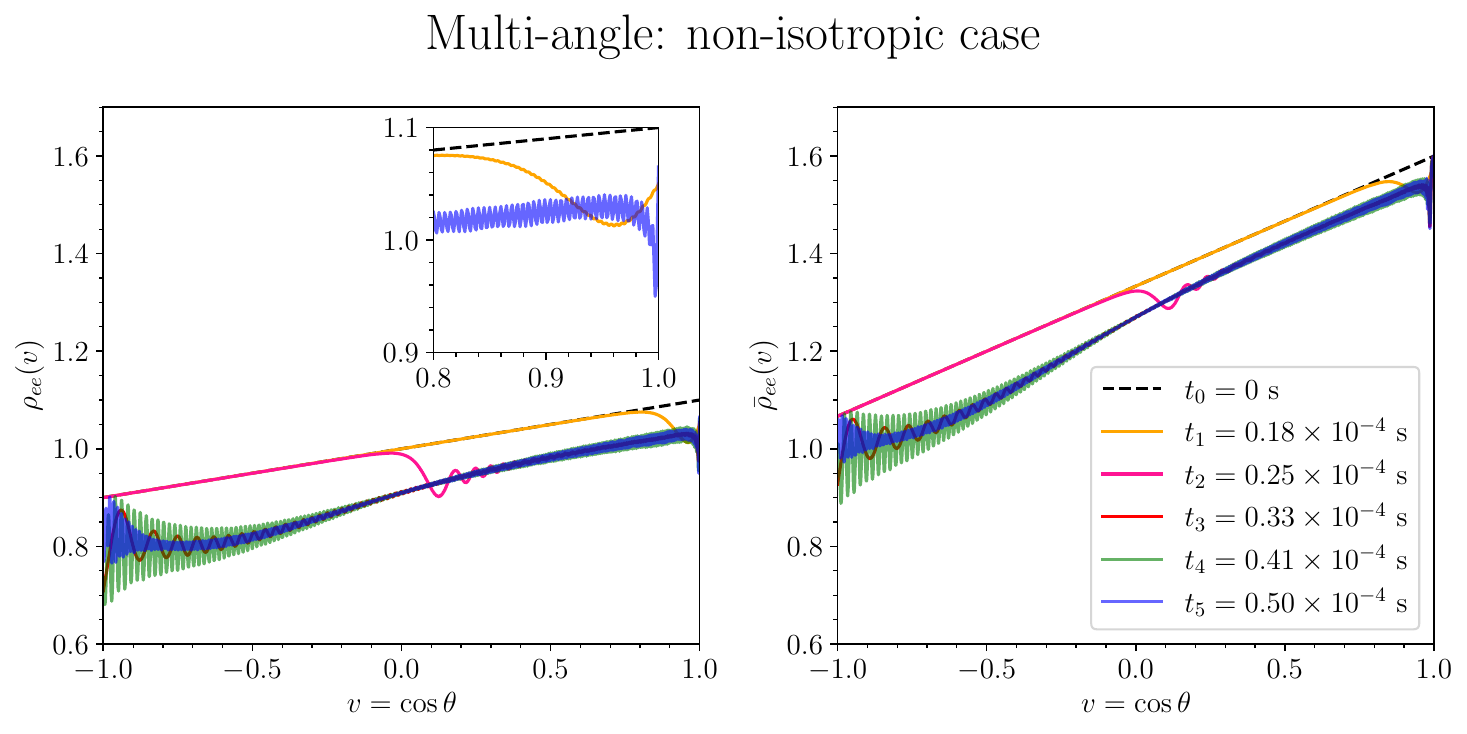}
\end{center}
\caption{Snapshots of the angular distributions of $\nu_e$ (left panel) and $\bar\nu_e$ (right panel) at the selected times $t_{1-5}$ for the MA solution for the non-isotropic neutrino gas in Fig.~\ref{fig:10a}. Due to the angular distributions of neutrinos being non-isotropic initially (dashed), different angular modes meet the MNR condition at different times. At first, only forward modes ($ v\sim 1$) evolve as shown by the orange curve in the inset. At later times the instability spreads across all other angle bins as seen in the green and blue curves.
}
\label{fig:11a}
\end{figure}
Figure~\ref{fig:11a} shows  snapshots of the  angular distributions of $\nu_e$ and $\bar\nu_e$ at the time snapshots $t_{1-5}$. The initial non-isotropic distributions are plotted with black dashed lines. At $t_1$, when the bin $v=1$ meets the MNR condition (see left panel of Fig.~\ref{fig:10a}), forward and backward angular bins start behaving differently. Notice that, between $t_1$ and $t_2$, only forward $v$ modes evolve (see orange and pink curves), and later, between $t_2$ and $t_3$, both forward and backward angular modes evolve. After all the angular bins have undergone  MNR, the system reaches a steady state at $t_3$. At $t \gtrsim t_4$ the system is highly non-linear and fine-grained angular structure arises.

\begin{figure*}[t!]
\includegraphics[width=0.99\textwidth]{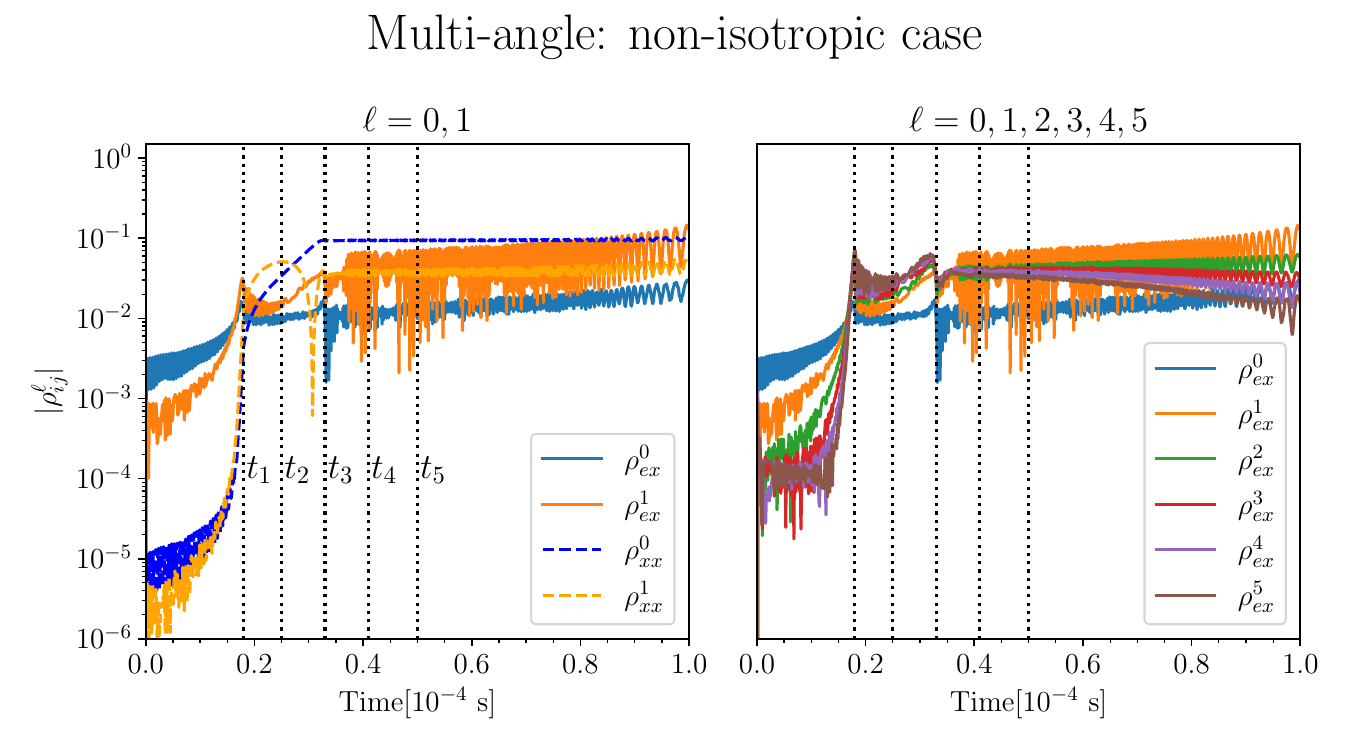}
\caption{
Temporal evolution of the Legendre coefficients of  $\rho_{ij}$ for the non-isotropic, MA neutrino gas in Figs.~\ref{fig:10a} and~\ref{fig:11a}. \textit{Left: } Time evolution of the norm of $\rho_{ex}^{0},\rho_{xx}^{0}, \rho_{ex}^{1}$ and $\rho_{xx}^{1}$. Unlike the isotropic MA system  in Fig.~\ref{fig:9a}, the components $\rho_{ex}^1$ (orange) and $\rho_{xx}^1$ (dashed yellow) are of $\mathcal{O}(10^{-3})$ and $\mathcal{O}(10^{-5})$, respectively, at  $t_0=0$. At $t_1$, when the first angular bin meets the MNR condition (see Fig.~\ref{fig:10a}), both $\ell=0$ and $\ell=1$ modes are comparable to each other in magnitude and thus already coupled to each other in the EOMs. \textit{Right: } Temporal evolution of the first six $\ell$-modes. Modes with $\ell>1$ are of the same order of magnitude already when the system enters the MNR at $t_1$. This is a crucial difference compared to the isotropic system, whose large $\ell$ modes are zero and grow during the MNR dynamically.}
\label{fig:12a}
\end{figure*}
Figure~\ref{fig:12a} displays the time evolution of the monopole, dipole, and high-$\ell$ modes for the neutrino density matrices. At  $t_1$, the total potential of the $v=1$ bin (black curve in the left panel of Fig.~\ref{fig:10a}) crosses the zero line, i.e.~meets the MNR condition. At $t_1$, all $\ell$ modes are comparable to each other in magnitude, as visible from the right panel of Fig.~\ref{fig:12a}. A crucial difference with respect to the isotropic case (Fig.~\ref{fig:9a}) is that larger $\ell>0$ modes are not negligible to begin with. As a result, high $\ell$ modes  couple to each other, preventing the oscillated potential $H_{ee}^{\nu\nu}(v)-H_{xx}^{\nu\nu}(v)+\lambda$ from staying in resonance. Between $t_1$ and $t_3$, the trend continues more or less unchanged. During  $\Delta t = t_3-t_1$,  all angular bins meet the MNR condition. After $t_3$, the last bin $v=-1$ (blue curve in left panel Fig.~\ref{fig:10a}) undergoes MNR, and the system reaches a steady state. For  $t \gtrsim t_4$, the $\ell>1$ modes are equal to  $\sim 10^{-2}$  until the end of the evolution at $10^{-4}$~s. This is reflected in the overall shape of the angular distributions at the $t_3$, $t_4$ and $t_5$, as visible from the red, green and blue curves in Fig.~\ref{fig:11a} which overlap with each other.

\section{Matter-neutrino resonance in  neutron star merger remnants}
\label{sec:BHR}

In order to appreciate the differences in the MNR flavor conversion physics between the SA and MA cases, we have used a simple idealized neutrino gas in the previous sections. In this section, we apply our findings to the BH accretion disk resulting from the binary neutron star post-merger phase.

\begin{figure*}
\begin{center}
\includegraphics[width=0.90\textwidth]{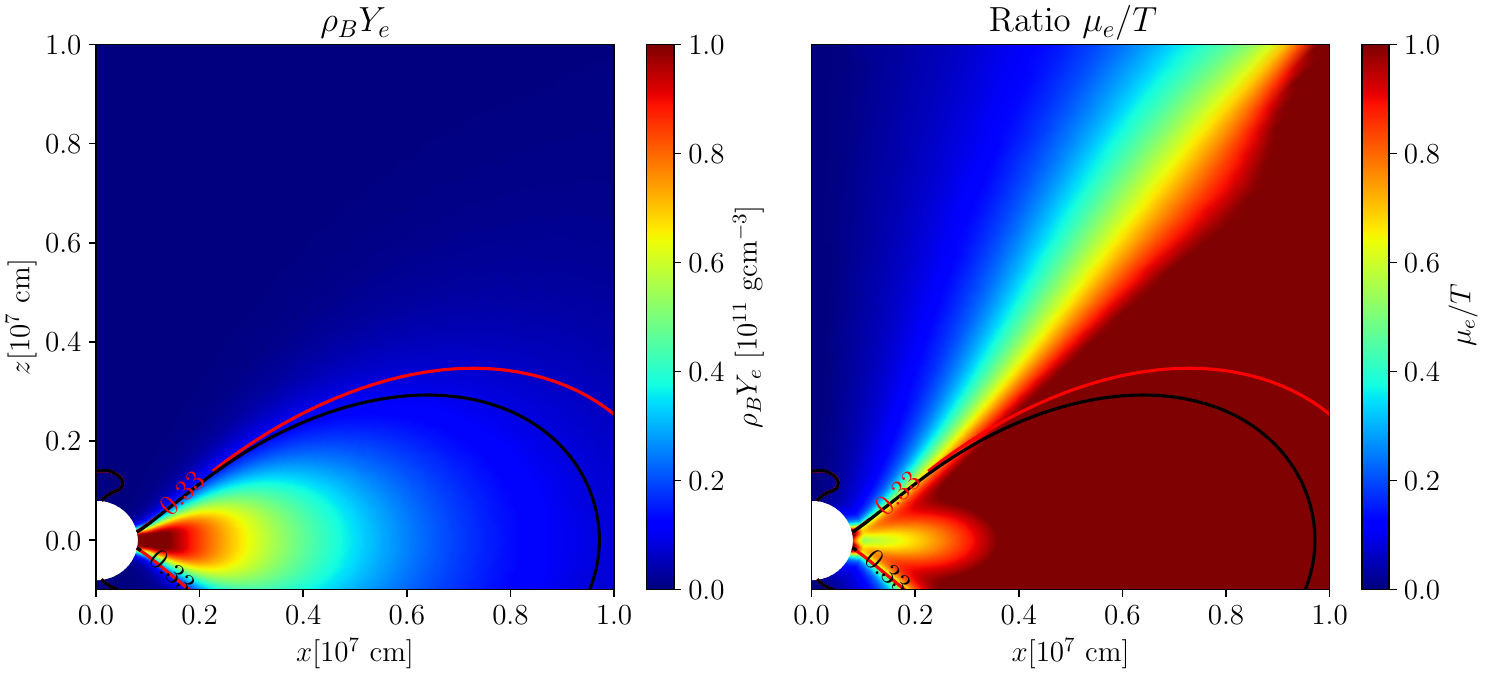}
\includegraphics[width=0.90\textwidth]{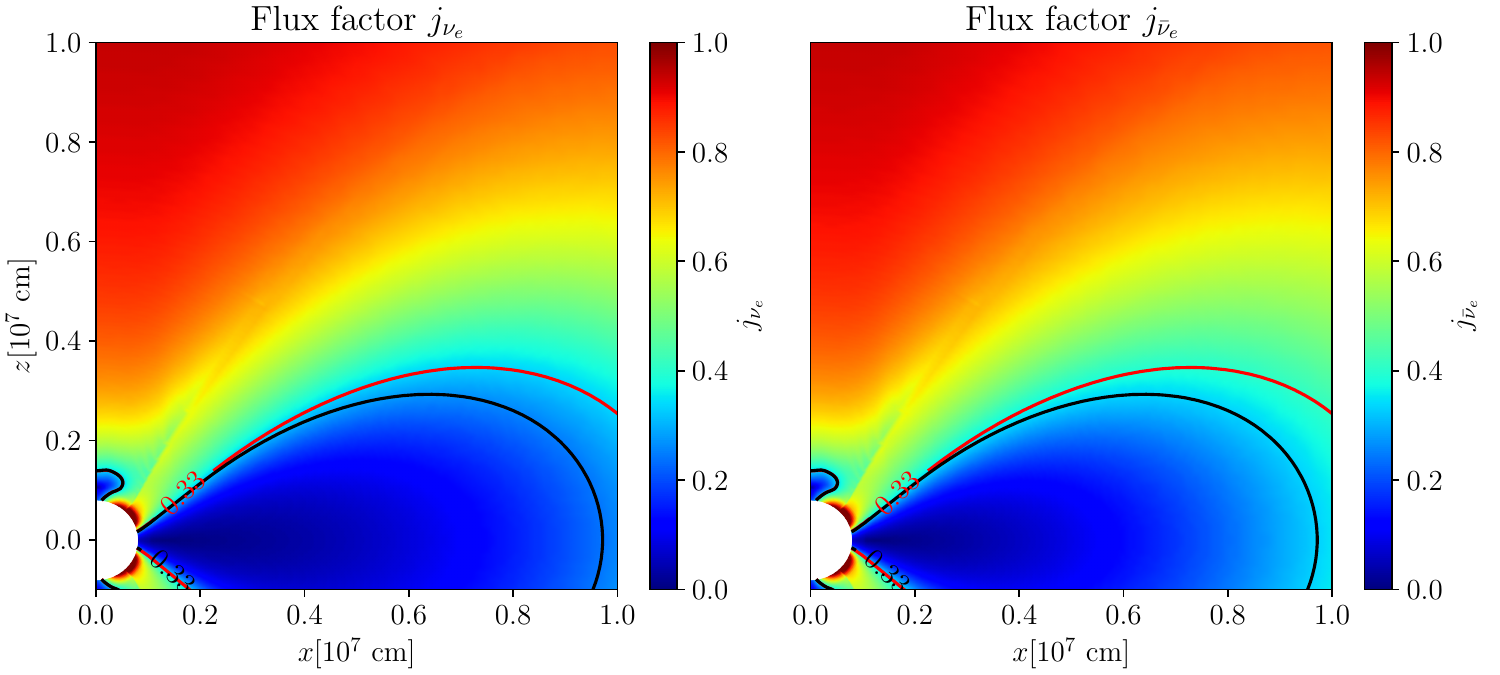}
\end{center}
\caption{Characteristic properties of the BH torus remnant at $15$~ms in  the $x$--$z$ plane.  \textit{Top: } Product of the baryon density $\rho_{B}$ and electron fraction $Y_e$ (i.e., electron density, left) and the ratio of the chemical potentials and temperature (right). Non-degenerate regions roughly correspond to locations where  $\mu_e/T \lesssim 1$. \textit{Bottom:} Electron neutrino (left) and antineutrino (right) flux factors as defined in Eq.~\ref{eq:fluxfac}. The solid lines represent the locations where the $\jnue = 1/3$ (red) and $\jnueb = 1/3$ (black)---inside these regions, $\nu_e$'s and $\bar\nu_e$'s are trapped and outside these regions, they are approximately in the free-streaming regime. 
}
\label{fig:14a}
\end{figure*}
\subsection{Neutron star merger remnant model}
We adopt the 2D hydrodynamical simulation output carried out in Ref.~\cite{Just_2015}. In particular, we consider the BH accretion torus model M3A8m3a5. This simulation employs neutrino moment transport and subgrid viscosity. An idealized equilibrium torus of $0.3\ M_\odot$ is simulated in the surroundings of a central BH of $3\ M_\odot$, with a dimensionless BH spin parameter of $0.8$.
The newly born accretion torus starts to lose mass while accreting on the BH. Neutrino cooling is not efficient in the first $\mathcal{O}(10)$~ms and the environment is optically thick. Then, as the density drops, neutrino cooling starts to balance viscous heating. The neutrino production rate decreases as the mass and density of the torus decrease until neutrino cooling becomes inefficient and the torus enters a phase dominated by advection, with viscous heating driving the expansion of the torus and launching the outflows. 

For simplicity and illustrative purposes, we focus on the disk properties at $15$~ms after the merger. Figure~\ref{fig:14a} shows the following physical quantities in the $x$--$z$ plane: the product between the baryon mass density $\rho_B$ and the electron fraction $Y_e$ (i.e., the electron density), the ratio of the electron potential $\mu_e$ and the electron temperature $T$, and the flux factors of neutrinos and antineutrinos, $j_{\nu_{e}}$ and $j_{\bar{\nu}_{e}}$. We can see that the region in the proximity of the polar axis is less dense with respect to the disk one. Moreover, $\mu_e/T \lesssim 1$ in a large portion of the polar region.

At a given location above the disk, the energy-integrated neutrino number flux vector $({\vec{F}_{\nu_\alpha}})$ and the neutrino number density $(n_{\nu_\alpha})$ are extracted from the hydrodynamical simulation (model M3A8m3a5) to compute the location of the electron neutrino and antineutrino decoupling surfaces. To this purpose, we require that the flux factors in the absence of flavor conversion are~\cite{Wu:2017drk}
\begin{eqnarray}
\label{eq:fluxfac}
    j_{\nu_e, \bar{\nu}_e} \equiv \frac{ | {\vec{F}_{\nu_e, \bar{\nu}_e }}| }{ n_{\nu_e, \bar{\nu}_e}} = \frac{1}{3} \ .
\end{eqnarray}
If the flux factor is small, then $\nu_e$'s or $\bar\nu_e$'s are expected to be strongly coupled to matter. On the contrary, for large values of the flux factor, i.e., $\mathcal{O}(j_{\nu_e, \bar{\nu}_e}) \sim 1$, $\nu_e$'s or $\bar\nu_e$'s are approximately in the free streaming regime. 
The red and black solid lines in Fig.~\ref{fig:14a}  are the locations where the flux factors equal $1/3$.
Note that, while   (anti)neutrinos are distributed according to an energy spectrum in model M3A8m3a5,  for simplicity we consider mono-energetic neutrinos as outlined in Sec.~\ref{sec:eoms}.

\subsection{Modeling of the neutrino angular distributions}

The BH torus simulation does not provide the angular distributions of $\nu_{e}$'s and $\bar{\nu}_{e}$'s. Hence, we
set them arbitrarily at the location above the disk where we start to evolve the EOM. We adopt two parametrizations for the angular distributions.

In the  case of isotropic  angular distributions, we assume
\begin{eqnarray}\label{eq:isorho}
    \rho_{ee}(v)=\frac{1}{2} \ \mathrm{and} \ \bar{\rho}_{ee}(v)=\frac{\alpha}{2} \ ,     
\end{eqnarray}
where $\alpha = n_{\bar{\nu}_e}/n_{\nu_e}$ is extracted from our benchmark BH torus model. 
\begin{figure}[t!]
\begin{center}
\includegraphics[width=0.49\textwidth]{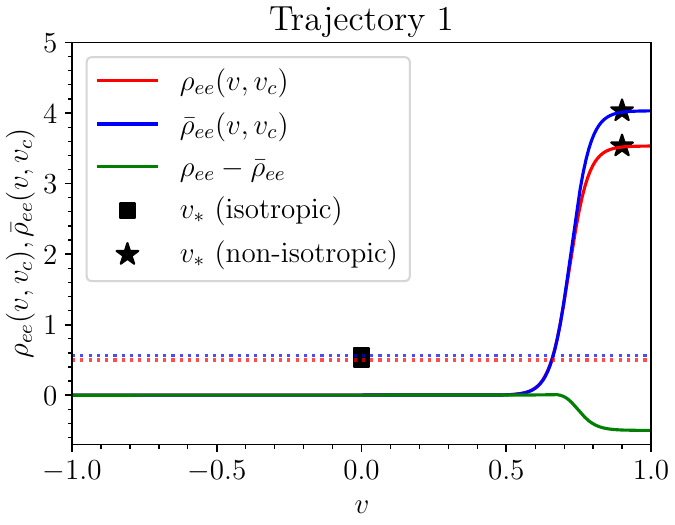}
\includegraphics[width=0.49\textwidth]{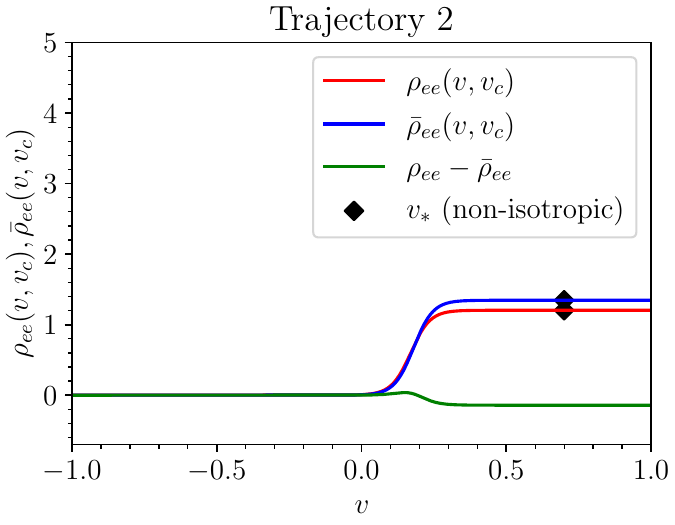}
\end{center}
\caption{\textit{Left:} Initial angular distributions (parametrized as in Eq.~\ref{eq:sigmoid}) of $\nu_e$ (solid red) and $\bar\nu_e$ (solid blue) and their difference (solid green) at $x_0=0.3\times 10^{6}$~cm, $z_0=0.5\times 10^{7}$~cm (cf.~star marker in Fig.~\ref{fig:15a}). The flux factors at this location are $\jnue=0.853$ and $\jnueb=0.855$, while the number densities are such that $\alpha \approx 1.11$. For comparison, the dotted lines represent the isotropic distributions (Eq.~\ref{eq:isorho}) for $\nu_e$'s  (red) and $\bar\nu_e$'s  (blue). \textit{Right: } Same as the left panel, but at $x_0=0.4\times 10^{6}$~cm, $z_0=0.2\times 10^{7}$~cm (diamond marker in Fig.~\ref{fig:15a}); since this location is closer to the central compact object, the flux factors are smaller ($\jnue=0.583$ and $\jnueb=0.588$). As a result, the angular distributions are less forward-peaked than the ones in the left panel. The number densities are such that $\alpha \approx 1.10$. Each marker (star, diamond and square) shows $v_*$ (for the cases shown in Fig.~\ref{fig:15a}) which corresponds to the angular bin where (anti)neutrinos start to plateau. In Table~\ref{tab:params} we summarize the parameters used for each case.
}
\label{fig:17b}
\end{figure}
\begin{figure}[t!]
\begin{center}
\includegraphics[width=0.60\textwidth]{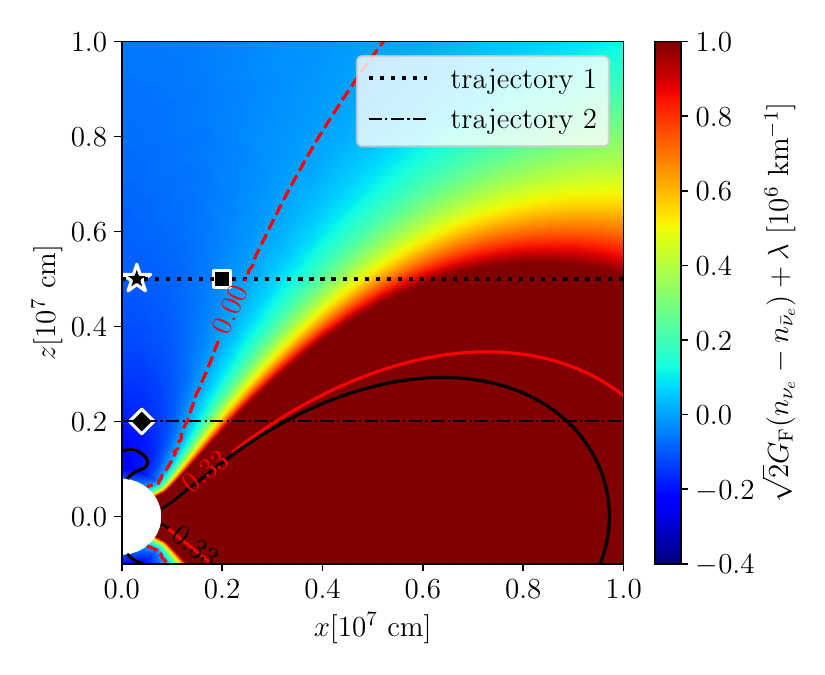}
\end{center}
\caption{Isocontours of the total potential [$\sqrt{2}\GF(n_{\nu_e}-n_{\bar{\nu}_e}) + \lambda$] in the $x$--$z$ plane in the SA (or angle-integrated MA isotropic) case.   The red and black solid lines represent the decoupling surfaces of neutrinos and antineutrinos, respectively. The red dashed line marks the loci where the neutrino potential $\sqrt{2}\GF(n_{\nu_e}-n_{\bar{\nu}_e})$ and the matter potential $\lambda$ cancel each other (Eq.~\ref{rescond}). The black dotted line represents a selected trajectory with $z=0.5\times 10^{7}$~cm (trajectory 1), and the black dash-dotted line is another trajectory with $z=0.2\times 10^{7}$~cm (trajectory 2) along which we solve the EOMs. The square, star and diamond markers show the loci at which the angular distributions are modeled.
}
\label{fig:15a}
\end{figure}
For the scenario with anisotropic angular distributions, we parametrize the latter with two independent variables, which can be uniquely determined given the first two neutrino moments obtained from the simulation data~\cite{Murchikova:2017zsy, Cernohorsky:1994yg, Richers:2022dqa}. We use a sigmoid function:
\begin{eqnarray}
    \rho_{ee}(v,v_c) = C\frac{1}{1+e^{-\beta(v-v_c)}} \ \ \mathrm{and}\ \ 
    \bar{\rho}_{ee}(v,\bar{v}_c) = \bar{C}\frac{\alpha}{1+e^{-\beta (v-\bar{v}_c)}} \ , 
\label{eq:sigmoid}
\end{eqnarray}
where the constants $C$ and $\bar{C}$ ensure that $\int dv\rho_{ee}(v)=1$ and $\int dv\bar{\rho}_{ee}(v)=\alpha$, 
respectively. The parameters $v_c$ and  $\bar{v}_c$ shift the distributions towards left or right in $v$ (the closer $v_c$ and $\bar{v}_c$ are to $1$ the more forward-peaked the distributions are), while the parameter $\beta= 30$ in the exponent softens the edges of the angular distributions; the larger $\beta$  the closer the angular distributions resemble top-hat distributions. Varying the value of $v_c$ ($\bar{v}_c$), we can match any neutrino (antineutrino) flux factor (Eq.~\ref{eq:fluxfac}) between $0$ and $1$; notice that since for example the flux factor for $\nu_e$ is the ratio between $\int dv v \rho_{ee}$ and $\int dv \rho_{ee}$ (similar for antineutrinos), the constants $C$, $\bar{C}$ and $\alpha$ drop out and the only tunable parameter left is $v_c$ ($\bar{v}_c$). In the limit $v_c\approx 1$, the flux factor is maximum ($\jnue\approx 1$), while for $v_c\approx -1$ the flux factor is minimum ($\jnue\approx 0$). Intuitively, these two extremes correspond to the cases of a Delta distribution function in the forward direction and a completely isotropic distribution, respectively.

We define $v_*$ as the first angular bin at which the (anti)neutrino angular distributions reach a plateau. Note that the value of $v_*$ is approximately the same for both neutrinos and antineutrinos because the distributions peak in the same angular region. This is a result of the (anti)neutrino flux factors being close to each other in magnitude (see bottom panels of Fig.~\ref{fig:14a}). If the angular distributions are isotropic, we consider $v_* = 0$.

Figure~\ref{fig:17b} shows the isotropic distributions for the square (dotted lines in the left panel), the non-isotropic angular distributions in Eq.~\ref{eq:sigmoid} for the star (solid lines in the left panel) and diamond (solid lines in the right panel) marker locations indicated in Fig.~\ref{fig:15a}. For the non-isotropic distributions in the left panel of Fig.~\ref{fig:17b} (trajectory 1), we use $v_c=0.717$ and $\bar{v}_c=0.723$ given that the correspondent flux factors are $\jnue=0.853$ and $\jnueb=0.855$ (see bottom panels of Fig.~\ref{fig:14a}). On the other hand,  in the right panel of Fig.~\ref{fig:17b}, we show the angular distributions extracted in correspondence of the diamond marker in Fig.~\ref{fig:15a} (trajectory 2), with $v_c=0.17$ and $\bar{v}_c=0.18$;  the flux factors at this location are  $\jnue=0.583$ and $\jnueb=0.588$. Table~\ref{tab:params} provides a summary of these parameters ordered by case. For the case of the star marker (trajectory 1, non-isotropic), we have $v_*=0.9$. For the diamond marker (trajectory 2, non-isotropic), we have  $v_*=0.7$. 
(Note that the star, diamond, and square markers are also used in Fig.~\ref{fig:15a} to indicate the   location above the merger disk whose  angular distributions  are used as initial conditions to evolve the EOMs.) 
\begin{table}[ht]
 \caption{Parameters for the angular distributions shown in Fig.~\ref{fig:17b} and their location above the torus (see Fig.~\ref{fig:15a}).}
    \label{tab:params}
    \centering
    \begin{tabular*}{\columnwidth}{@{\extracolsep{\fill}}lllllll}
    \hline\hline
     Marker   & $(x_0,z_0)[10^7 \ \mathrm{cm}]$ & $(\jnue,\jnueb)$ & $(v_c,\bar{v}_c)$ & $v_*$   & $\alpha$ & Label \\
     \hline
     Square   & $(0.20, 0.50)$  & -- & -- & 0.0 & 1.12 & traj. 1 (iso.)   \\
     Star     & $(0.03, 0.50)$ & $(0.853,0.855)$ & $(0.717,0.723)$ & 0.9 & 1.11 & traj. 1 (non-iso.) \\
     Diamond  & $(0.04, 0.20)$ & $(0.583,0.588)$ & $(0.170,0.180)$ & 0.7 & 1.10 & traj. 2 (non-iso.) \\
     \hline
    \end{tabular*}
    \vskip12pt
\end{table}

The angular distributions in the left panel of Fig.~\ref{fig:17b} do not cross each other, implying the absence of an electron lepton number crossing necessary for fast flavor evolution~\cite{Izaguirre:2016gsx, Morinaga:2021vmc, Padilla-Gay:2021haz}. On the other hand,  the angular distributions of $\nu_e$ and $\bar\nu_e$  in the right panel of Fig.~\ref{fig:17b} exhibit a small angular crossing (see solid green line); however, we have checked that the system is stable under fast flavor instabilities. Moreover, the initial angular distributions for trajectory 1 and trajectory 2 are such that we have an overall excess of $\bar{\nu}_{e}$ over $\nu_{e}$, as well as an overall excess of (anti)neutrinos along the forward ($v=1$) direction; antineutrinos are, however, not only more abundant but also more forward-peaked, resembling the flux factor distributions.

\subsection{Characterizing the matter and neutrino-neutrino interaction potentials}

To investigate the neutrino flavor conversion physics, we compute the matter potential  from Eq.~\ref{eq:lambda}, while the neutrino self-interaction potential  follows from Eq.~\ref{eq:hnunu}, hence for example
\begin{eqnarray}
    H_{\nu\nu}^{ee}(v,t_0) = \sqrt{2}\GF n_{\nu_e} \int dv^\prime \left[\rho_{ee}(v^\prime,t_0)-\alpha \bar{\rho}_{ee}(v^\prime,t_0)\right][1-v v^\prime]\ .
\end{eqnarray}
We rely on the output data of the model M3A8m3a5 to compute the diagonal elements of $H_{\nu\nu}(v,t_0)$ as illustrated above. 

Figure~\ref{fig:15a} shows isocontours of the total potential, $\sqrt{2}\GF(n_{\nu_e}-n_{\bar{\nu}_e})+\lambda$, in the MA isotropic (or SA) approximation and the $x$--$z$ plane. The red (blue) regions in the $x$--$z$ plane indicate the locations where the total potential is positive (negative). This means that the MNR condition (Eq.~\ref{rescond}) is satisfied along the isocontour which we show through the red dashed line.

In order to solve the EOMs, we pick trajectories 1 and 2, both crossing the red dashed line in Fig.~\ref{fig:15a}, i.e.~the black dotted and dash-dotted lines along $z=0.5\times 10^{7}$~cm and $z=0.2\times 10^{7}$~cm, respectively. At each point along these trajectories, we compute the matter and neutrino-neutrino potentials relying on the outputs of our BH disk model. We construct the initial angular distributions at the initial $(x_0, z_0)$ locations above the BH torus (see Fig.~\ref{fig:17b}) as illustrated in the previous subsection and then let the angular distributions evolve due to flavor conversion while neglecting neutrino advection and non-forward collisions as discussed in Sec.~\ref{sec:eoms}.

\begin{figure}[t!]
\begin{center}
\includegraphics[width=0.49\textwidth]{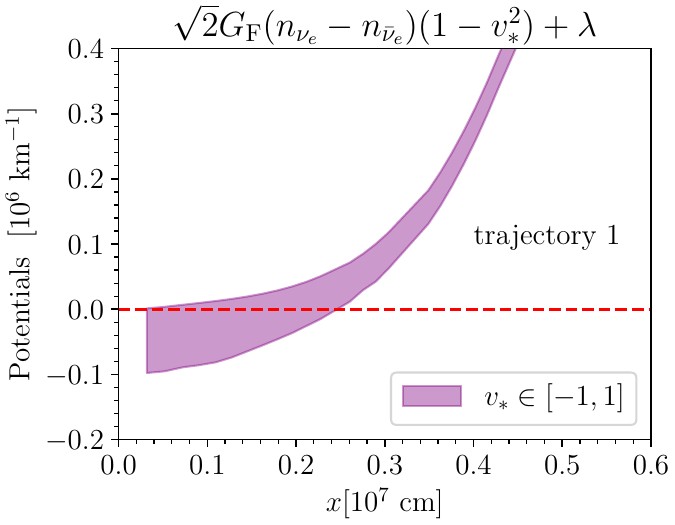}
\includegraphics[width=0.49\textwidth]{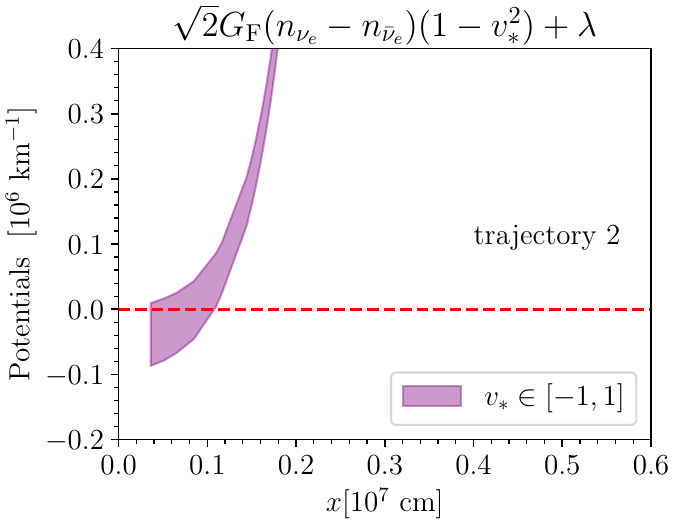}
\end{center}
\caption{\textit{Left:} Matter and neutrino-neutrino potentials along trajectory 1 as a function of $x$. The spatial domain covers a portion of the horizontal dotted line in Fig.~\ref{fig:15a}. The MNR condition is fulfilled within $\Delta x=10^6$ cm from positive to negative values of the total potential. The purple-shaded area corresponds to the maximum variation of the total potential. To guide the eye, the red dashed line shows the zero value for the total potential. \textit{Right:} Same as the left panel, but for trajectory 2  (dash-dotted line in Fig.~\ref{fig:15a}). Given the much closer location to the central BH at which trajectory 2 starts, the MNR condition can be fulfilled at smaller values of $x$. 
}
\label{fig:17a}
\end{figure}

The ``average'' potential felt by neutrinos should be different for the isotropic and anisotropic MA configurations. In order to take this into account when exploring where the MNR condition is expected to take place,  we introduce modified total potentials in Fig.~\ref{fig:17a}. For illustration, we consider two extreme angular distributions: an isotropic angular distribution and a Delta function [$\delta(v_* - v)$]. Since in the MA framework, the MNR condition depends on $v$, these two angular distributions should meet the MNR condition at different $x$ and  $v$ on average. For the isotropic case, the MNR condition is given by $\sqrt{2}\GF(n_{\nu_{e}}-n_{\bar{\nu}_{e}})+\lambda=0$;  for the Delta distribution, the MNR condition is on average given by $\sqrt{2}\GF(n_{\nu_{e}}-n_{\bar{\nu}_{e}})(1-v_*^2)+\lambda=0$.

Figure~\ref{fig:17a} represents the sum of the matter and neutrino self-interaction potentials as functions of $x$  for trajectory 1 (left panel) and trajectory 2 (right panel), see also  Fig.~\ref{fig:15a}. The purple-shaded bands in Fig.~\ref{fig:17a} represent the variation of the total potential. The lower limit of the band corresponds to the case of isotropic angular distributions ($v_*^2=0$). The upper limit corresponds to the case where the first term in the modified total potential is zero ($v_*^2=1$) and no MNR is expected because such potential is always positive. Intermediate values of $v_*$ shift the modified potential along $x$, which means that the MNR could occur close to the central compact object depending on how forward-peaked the angular distributions are.

For the initial conditions summarized in Table~\ref{tab:params}, the interval  $\Delta x = 10^{6}$ cm is sufficient for all cases to fulfill the MNR condition for a large number of angular bins. For all cases shown in Table~\ref{tab:params}, the total potentials are negative at their initial locations $(x_0,z_0)$ but become positive as one moves towards larger values of $x$, thereby fulfilling the MNR conditions around $x=0.1$--$0.3\times 10^{7}$~cm depending on the specific value of $v_*$.

We also investigate the MNR conditions along trajectory 2 (see dash-dotted line in Fig.~\ref{fig:15a}), which lies closer to the neutrino decoupling regions and has less forward-peaked distributions with respect to the ones of trajectory 1, while still maintaining a certain level of anisotropy. The initial angular distributions of trajectory 2 are shown in the right panel of Fig.~\ref{fig:17b} with its parameters summarized in Table~\ref{tab:params}.

\subsection{Neutrino flavor evolution along selected trajectories}\label{sec:BH_evolution}

\begin{figure*}[t!]
\includegraphics[width=0.94\textwidth]{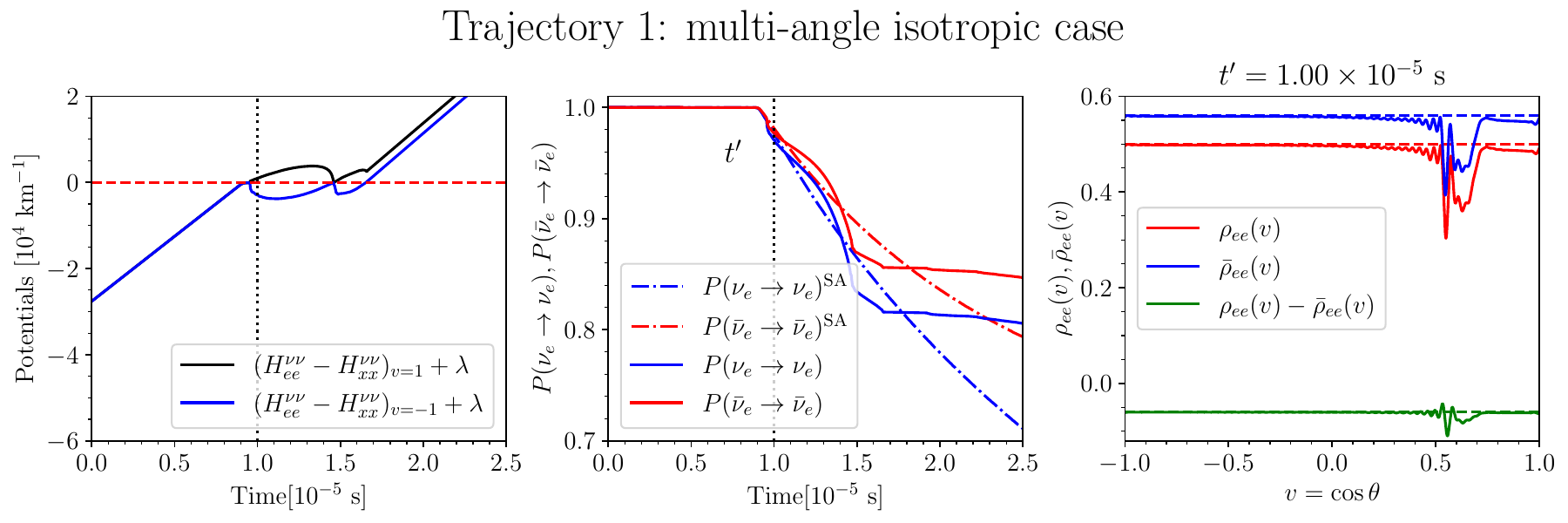}
\includegraphics[width=0.94\textwidth]{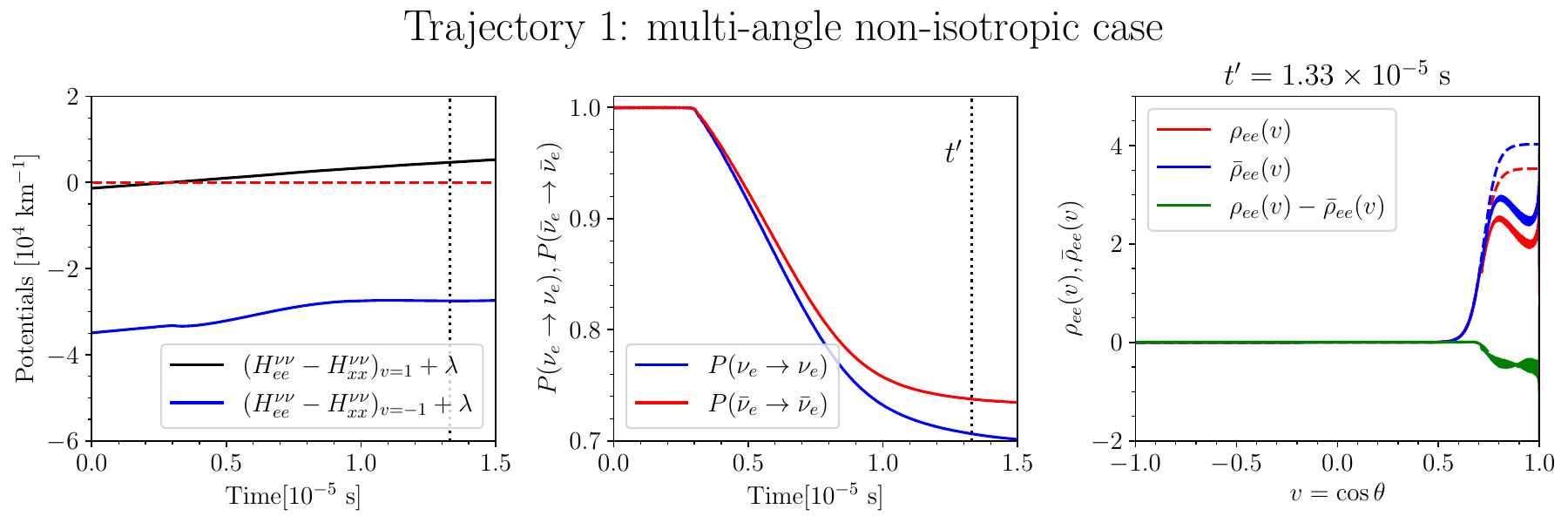}
\includegraphics[width=0.94\textwidth]{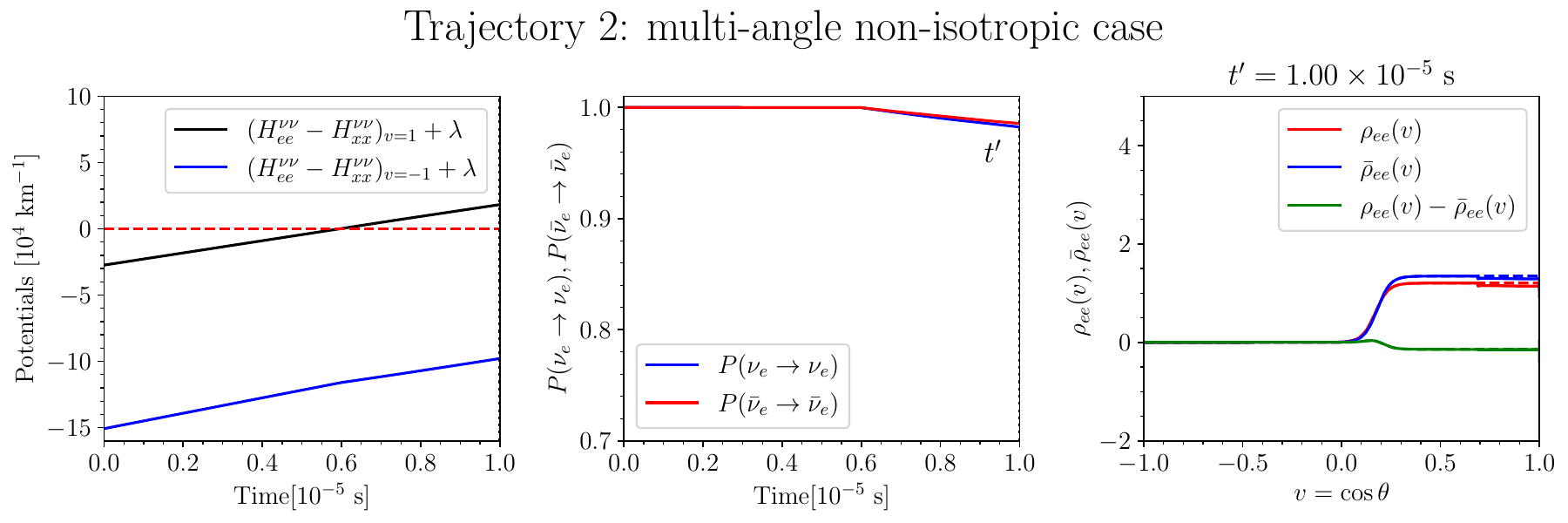}
\caption{Neutrino flavor conversion along trajectory 1, with isotropic (top row, Eq.~\ref{eq:isorho}) and non-isotropic (middle row, Eq.~\ref{eq:sigmoid}) angular distributions, as well as for non-isotropic angular distributions and trajectory 2 (bottom row). The flavor evolution starts at the $x_0$ location marked in Fig.~\ref{fig:17a} and Table~\ref{tab:params}. The left panels show the oscillated total potentials. The middle panels show the temporal evolution of the survival probabilities for neutrinos and antineutrinos.  The right panels show snapshots of the angular distributions at the selected time $t^\prime$ marked by the vertical dotted line.
For the isotropic case (top panels), our results are very similar to the ones in Fig.~\ref{fig:6a} where a dipole mode grows during the partial MNR transition. For this case, we show the corresponding SA system (dash-dotted lines in the top middle panel). In the non-isotropic case (middle row), there is no MNR transition. In the bottom panels, we show the results for trajectory 2. Here, a large number of forward angular bins undergo MNR, and flavor conversion begins as the first forward bin is involved in the MNR. However, the angular distributions along trajectory 2 are less forward-peaked than the ones of trajectory 1, and minimal flavor conversion takes place. 
}
\label{fig:18a}
\end{figure*}

We now present the solution of the EOMs along the selected trajectories 1 and 2 shown in Fig.~\ref{fig:15a},  whose total potentials in the absence of flavor conversion are displayed in Fig.~\ref{fig:17a}. Animations of the temporal evolution of the characteristic quantities for the selected trajectories are provided as \href{https://sid.erda.dk/share_redirect/e2zTyjhG3B/index.html}{Supplemental Material}. Note that, because of the simplified setup adopted in this work, the solution of the EOMs depends on where along a given trajectory we choose to start the simulation, i.e.~on $(x_0,z_0)$ and the initial configuration corresponding to the square, star and diamond markers in Fig.~\ref{fig:15a} (see also Table~\ref{tab:params}). The initial location $(x_0,z_0)$ is chosen such that the total potential starts from a negative value that is not too far from zero. This means that we choose the initial location sufficiently close to the MNR condition so that our numerical routine can cross the MNR within a reasonable amount of CPU time. Although the exact location $(x_0,z_0)$ is important for extracting the flux factors (see Table~\ref{tab:params}), small displacements do not significantly change our angular distributions. Moreover, since there are no ELN angular crossings, the system is stable between $(x_0,z_0)$  and the location where the first angular bin ($v=1$) meets the MNR condition.

We  adjust the vacuum mixing angle to $\theta_V=0.3$ (instead of $\theta_V=0.15$; see Sec.~\ref{sec:mnr}). The reason is that the potentials from the BH disk simulation are such that a slightly larger $\theta_V$ is needed to see an MNR transition even in the SA scenario. Thus,  if we do not find  MNR  for this large $\theta_V$, it is not because of an inadequate choice of vacuum mixing parameters (see Sec.~\ref{sec:sa} and Appendix~\ref{sec:single}) but because of MA effects.  

The top and middle rows of Fig.~\ref{fig:18a}  show the flavor evolution of (anti)neutrinos in the isotropic (Eq.~\ref{eq:isorho}) and non-isotropic scenarios (Eq.~\ref{eq:sigmoid} as in Fig.~\ref{fig:17a}) along trajectory 1--black dotted line in Fig.~\ref{fig:15a}. These results are qualitatively very similar to the ones presented for the ideal neutrino gas in Sec.~\ref{sec:mnr}. The flavor outcome in the SA and MA-isotropic cases is qualitatively different, as visible from the middle panel in the first row of Fig.~\ref{fig:18a}. The development of fine angular structures is the result of the diffusion of flavor waves from large to small angular scales which hinders the MNR development.

For trajectory 1, the MA isotropic case shows the least amount of flavor conversion, while the MA non-isotropic case is prone to significant flavor conversion. This difference in the flavor outcome is because, for forward-peaked distributions, almost all neutrinos are clustered in a narrow angular region and undergo the MNR  roughly at the same time. In the isotropic case,  neutrinos are equally distributed across $-1<v<1$ and undergo MNR progressively (more slowly) as $H_{ee}^{\nu\nu}(v)-H_{xx}^{\nu\nu}(v)+\lambda$ crosses zero for any $v$. 

For trajectory 2, the bottom panels of Fig.~\ref{fig:18a} show the flavor outcome in the MA non-isotropic case. In trajectory 2, (anti)neutrinos almost completely populate the angular region with $0<v<1$, with no (anti)neutrinos in the backward direction. Similar to the isotropic case of trajectory 1, no significant flavor conversion is found.
For both trajectory 1 (isotropic) and trajectory 2 (non-isotropic), the survival probabilities of (anti)neutrinos remain close to $0.9$, unlike the non-isotropic case in trajectory 1, where survival probabilities are approximately equal to $0.7$ after $1.5\times10^{-5}$~s.

These results highlight the crucial differences between SA and MA solutions of the EOMs and suggest that the MNR may only be relevant far from the decoupling regions where the angular distributions are forward-peaked. Alternatively, the MNR is unlikely to lead to significant flavor conversion in regions where the (anti)neutrino distributions are isotropic.

\section{Conclusions}\label{sec:conclusions}
In neutron star merger remnants,  neutrinos could undergo the MNR, characterized by a cancellation of the matter and neutrino-neutrino interaction potentials. In this paper, we explore the flavor conversion outcome caused by the MNR, solving the EOMs in the SA and MA scenarios. In the MA case, we explore the flavor conversion physics for initially isotropic and non-isotropic angular distributions. Under the assumptions of spatial homogeneity and mono-energetic, ideal (anti)neutrino gas, we find a significant difference in the flavor outcome between the SA and MA cases. Intriguingly,  the MA calculation of neutrino flavor evolution in the presence of MNR results in the spontaneous breaking of isotropy for both neutrino mass orderings. 

We further investigate the MNR physics relying on the output of a 2D hydrodynamical simulation of a BH accretion disk, resulting from the post-merger phase of a neutron star merger. Focusing on a representative time snapshot, we solve the EOMs along two selected trajectories above the accretion disk. We find that, depending on whether we employ initially isotropic neutrino angular distributions or non-isotropic ones, the flavor outcome is qualitatively very different. In the former case, no significant flavor conversion is found after the MNR is partially achieved. In the latter case, significant flavor conversion is triggered by the MNR condition being fulfilled for some angular bins; this scenario on average tends to resemble what would happen in the SA approximation, despite the isotropy-breaking feature typical of the MA case. These findings suggest that the flavor outcome resulting from the MNR is sensitive to the shape of the angular distributions, where forward-peaked distributions lead to more substantial flavor conversion because the MNR condition is met for the most populated angular bins around the peak of the distribution almost simultaneously. This is not the case for initially isotropic neutrino distributions, where the MNR condition is met across all equally-populated bins over a wider spatial range.

Our preliminary findings highlight an intrinsic difference between the SA and MA solutions of the flavor evolution in the presence of the MNR. 
This work also suggests a strong directional dependence of the MNR physics in neutron star merger remnants and calls for the development of a self-consistent modeling of the neutrino angular distributions. 
We stress that our modeling of the MNR in the MA framework is simplistic as it does not include a self-consistent evolution of the neutrino angular distributions based on non-forward collisions with the medium and neutrino advection. Furthermore, the possible interplay between the MNR physics and  flavor conversion triggered by fast~\cite{Wu:2017qpc,Wu:2017drk,Just:2022flt,George:2020veu,Li:2021vqj,Fernandez:2022yyv} as well as  collisional instabilities~\cite{Johns:2021qby,Johns:2022yqy,Xiong:2022zqz, Padilla-Gay:2022wck} should be  explored.

\acknowledgments
We are grateful to Oliver Just for providing access to the output data of the neutron star merger remnant simulation adopted in this work and  Alexander Friedland for useful discussions. IPG acknowledges support from the U.S. Department of Energy under contract number DE-AC02-76SF00515. In Copenhagen, this project has received support from the Danmarks Frie Forskningsfond (Project No.~8049-00038B), the Villum Foundation (Project No.~13164), and the European Union (ERC, ANET, Project No.~101087058). Views and opinions expressed are those of the authors only and do not necessarily reflect those of the European Union or the European Research Council. Neither the European Union nor the granting authority can be held responsible for them.


\begin{appendix}

\section{Characteristic timescales of the matter-neutrino resonance in the single-angle approximation}\label{sec:single}

\begin{figure*}[t!]
\includegraphics[width=0.99\textwidth]{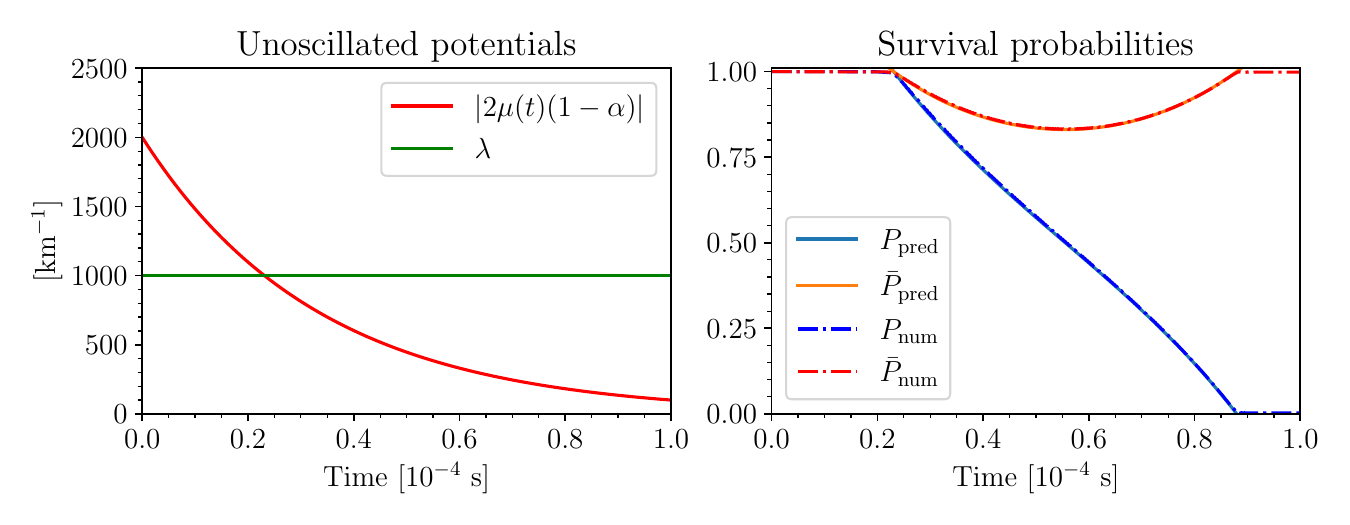}
\caption{\textit{Left:} Unoscillated self-interaction (red) and matter potentials (green) as functions of time. \textit{Right:} Survival probabilities of neutrinos (blue) and antineutrinos (red), obtained by solving the EOMs, as functions of time. The dashed lines are the predicted survival probabilities as shown in Eqs.~3-4 in Ref.~\cite{Malkus:2014iqa}; we can see that the numerical solution is in excellent agreement with the theoretical prediction. 
}
\label{fig:1a}
\end{figure*}

A clear feature of the  MNR in the SA approximation is the complete conversion of $\nu_e$, while $\bar{\nu}_e$ returns to its original configuration. In Fig.~\ref{fig:1a}, we show our numerical solution together with the analytical prediction for the survival probabilities. The duration of the single-angle MNR transition $\Delta t $ is well understood and can be estimated analytically. Reference~\cite{Malkus:2014iqa} derived the adiabaticity criterion from matching the timescale of the transition ($\delta T_1$) with the timescale set by the capacity of the system to change flavor  ($\delta T_2$). We estimate these timescales to validate our SA calculations. The duration of the transition is set by the timescale
\begin{eqnarray}\label{eq:deltaT1}
    \delta T_1 \approx \tau_{\lambda/\mu} \log{ \left[ {(1+\alpha)}/{(\alpha-1)}\right] } \ , 
\end{eqnarray}
where $\tau_{\lambda/\mu} $ is the effective timescale of the ratio of the potentials, namely
\begin{eqnarray}
    \tau_{\lambda/\mu} = \left| {d\log[\lambda/\mu(t)]}/{dt} \right|^{-1} \ .
\end{eqnarray}
Using the potentials in Eq.~\ref{eq:mulam}, it is straightforward to show that $\tau_{\lambda/\mu}=q^{-1}$ s, which  gives $\delta T_1 \simeq 6.49 \times 10^{-5}$ s. The agreement between this prediction and our numerical simulations can be seen in the upper panel of Fig.~\ref{fig:6a}, where the transition timescale is $\Delta t \simeq 6.50 \times 10^{-5}$ s and therefore  $\Delta t \approx \delta T_1$.

\begin{figure}[t!]
\begin{center}
\includegraphics[width=0.49\columnwidth]{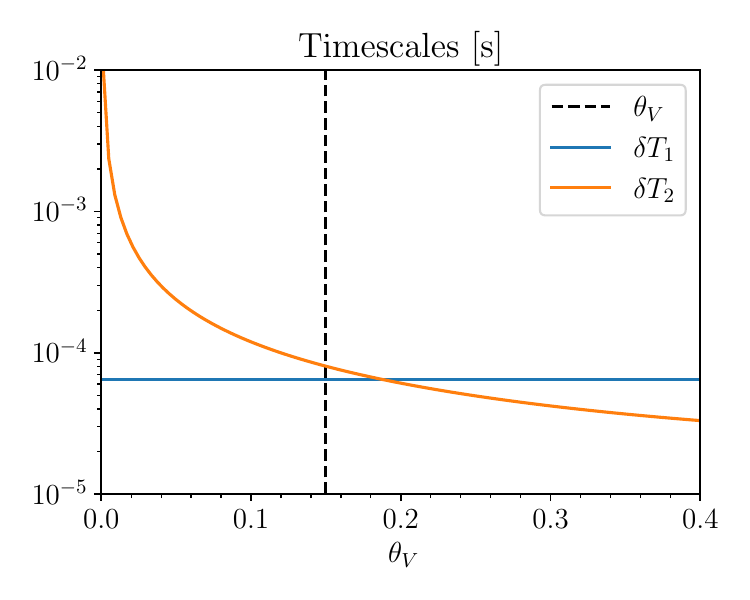}
\end{center}
\caption{Duration of the transition $\delta T_1$ and the timescale of the transition $\delta T_2$ as  functions of the mixing angle $\theta_V$. Timescales are comparable around our  $\theta_V=0.15$ marked through the dashed vertical line. Choosing $\theta_V\approx 0$ results in very different different timescales $\delta T_1$ and $\delta T_2$, thus preventing MNR transitions.}
\label{fig:3a}
\end{figure}

For MNR transitions to occur, the timescale $\delta T_1$ has to be comparable to the distance scale of the transition $\delta T_2$. The latter is given by
\begin{eqnarray}\label{eq:deltaT2}
    \delta T_2 \approx \frac{\alpha}{\omega\sin{2\theta_V}\langle\mathrm{Im}(\rho_{ex}+\alpha\bar{\rho}_{ex})\rangle} \ , 
\end{eqnarray}
where the term $\langle \cdot \rangle$ in the denominator is the average value of the imaginary component of the off-diagonal entries. For $\delta T_1$ and $\delta T_2$ to be comparable, $\langle \cdot \rangle$ must adjust to the vacuum oscillation parameters; clearly, for  $\omega\approx 0$ or $\theta_V \approx 0$, this requirement cannot be fulfilled. The mass hierarchy does not play an important role in the occurrence of MNR transitions, as it simply shifts the total potential by a negligible amount. However, the magnitudes of $\omega$ and $\theta_V$ set the timescale $\delta T_2$. In Fig.~\ref{fig:3a}, we show the timescales $\delta T_1$ (Eq.~\ref{eq:deltaT1}) as a function of the vacuum mixing angle $\theta_V$, while $\delta T_2$ (Eq.~\ref{eq:deltaT2}) is fixed by the choice of the potentials. The difference between both timescales is largest for very small $\theta_V$. The dashed line marks our choice with $\theta_V=0.15$, which results in comparable $\delta T_{1,2}$, so MNR transitions are possible. 

\begin{figure*}[t!]
\begin{center}
\includegraphics[width=0.80\textwidth]{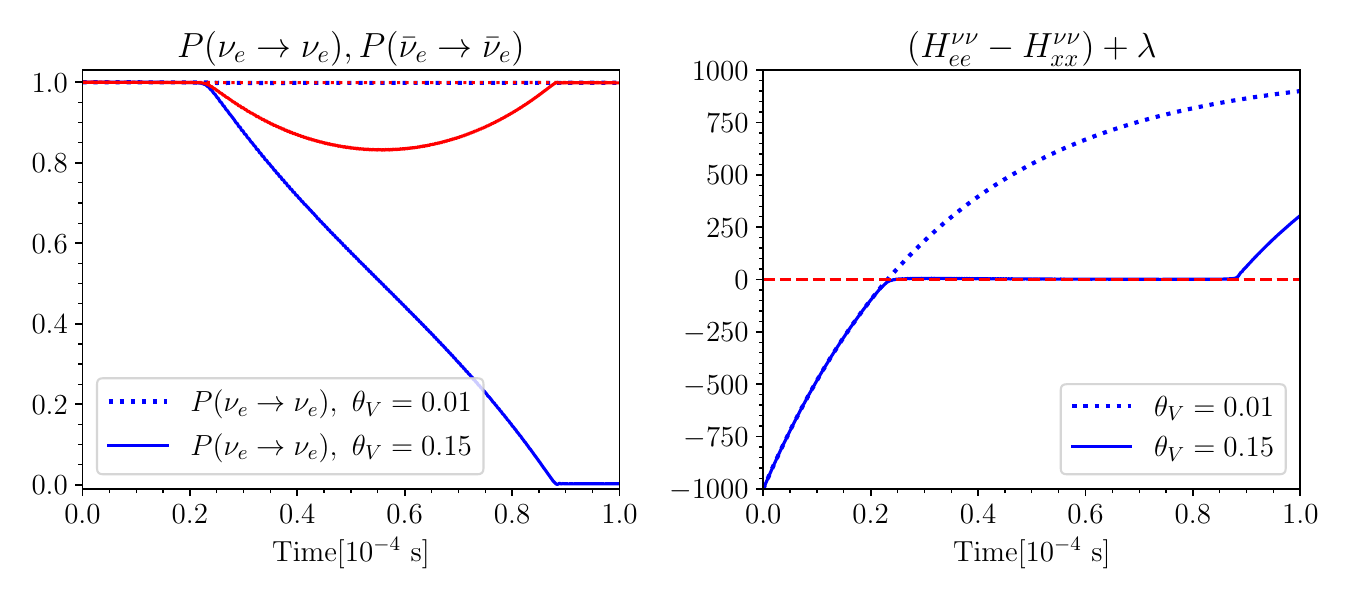}
\end{center}
\caption{Same as the upper panels of Fig.~\ref{fig:6a} but for two different values of the vacuum  mixing angle $\theta_V$. The survival probabilities (left panel) do not evolve for small $\theta_V=0.01$ (dashed) because the timescales $\delta T_1$ and $\delta T_2$ are too far apart (see also Fig.~\ref{fig:3a}) and the MNR transition is not possible. As a result, the total potential (right panel) for the system with $\theta_V=0.01$ (dashed) continues to increase without staying zero after $\sim 0.2\times 10^{-4}$ s.}
\label{fig:4a}
\end{figure*}

\begin{figure*}[t!]
\begin{center}
\includegraphics[width=0.80\textwidth]{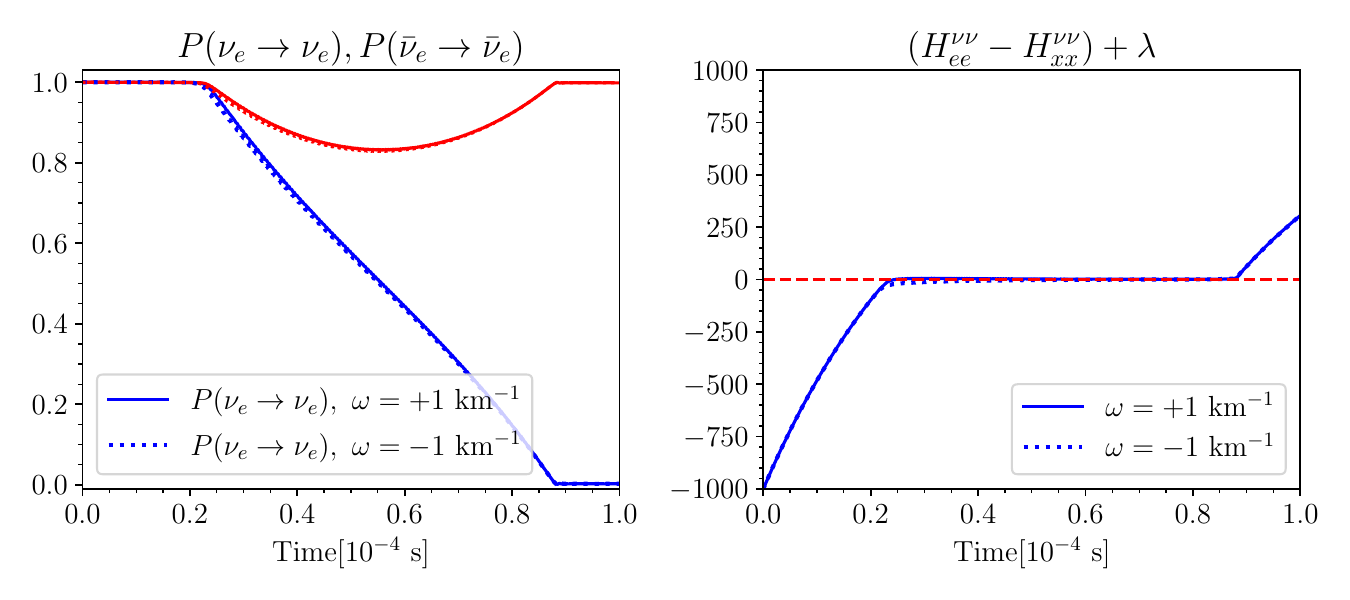}
\end{center}
\caption{Same as the upper panels of Fig.~\ref{fig:6a} but for normal ($\omega>0$) and inverted ($\omega<0$) mass orderings, shown in solid and dotted lines, respectively. In agreement with  Eqs.~3-4 of Ref.~\cite{Malkus:2014iqa}, both mass orderings display qualitatively similar dynamics and only small deviations are visible. The value of $\omega$ shifts the total potential (right panel) by a small amount around $t=0.2\times10^{-4}$~s and has no overall impact on the outcome.}
\label{fig:5a}
\end{figure*}

In Fig.~\ref{fig:4a}, we show the flavor evolution for $\theta_V=0.01$ and  $\theta_V=0.15$ (i.e., small and large mixing angles), where the former does not display MNR transitions while the latter does.   Figure~\ref{fig:5a} confirms that the mass ordering does not play an important role in the occurrence of MNR transitions as it simply shifts the total potential by a small amount. The magnitude of $\omega$, however, sets the timescale $\delta T_2$ together with $\theta_V$.

Although not shown here, we have performed simulations where the roles of $\mu$ and $\lambda$ are swapped [i.e.~$\lambda(t) \propto \exp(-q t)$ and time independent $\mu$]. In this case, the cancellation of the potentials is also possible. One would naively expect that an MNR would occur in this scenario as well. This is, however, not the case. The reason is that, if the roles of $\mu$ and $\lambda$ are swapped,  $\tau_{\lambda/\mu}=q=3\times 10^4$~s  (instead of $q^{-1}=0.33\times 10^{-5}$~s), which results in $\delta T_{1}$ and $\delta T_{2}$ being a few orders of magnitude apart. Our numerical results confirm the absence of MNR transitions for this scenario in agreement with Fig.~1(b) of Ref.~\cite{Malkus:2014iqa}.

\begin{figure*}[t!]
\begin{center}
\includegraphics[width=0.60\textwidth]{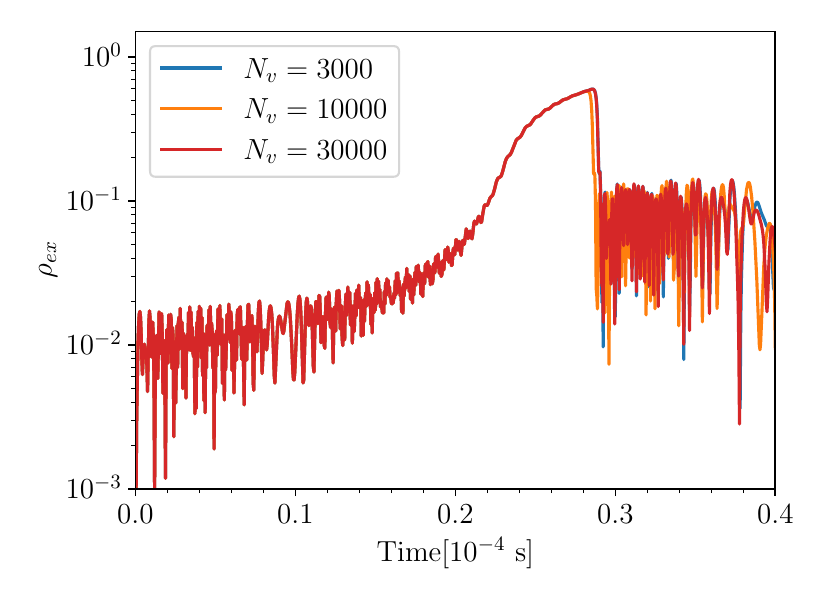}
\end{center}
\caption{Evolution of $\rho_{ex}$ for a similar isotropic system shown in the lower panels of Fig.~\ref{fig:6a} for different numbers of angular bins. We compare a low-resolution run with higher-resolution runs. Although the non-linear regime is affected by the number of angle bins, the general behavior and key features discussed in Sec.~\ref{sec:mnr} are preserved as the angular resolution is increased. The breaking of angular symmetry still occurs around $t\simeq 0.3\times10^{-4}$ s for the low and high-resolution runs.}
\label{fig:13a}
\end{figure*}

\section{Numerical convergence of the solution of the neutrino equations of motion}\label{sec:resolution}

To evolve the (anti)neutrino EOMs in time, we use the Runge-Kutta-Fehlberg(7,8) method with adaptive stepsize from the odeint library of Boost~\cite{BoostLibrary}. It is well-known that MA calculations can display spurious instabilities caused by a lack of resolution~\cite{Sarikas:2012ad, Morinaga:2018aug}. To ensure that our simulations are not prone to spurious instabilities, we adopt very high angular resolution throughout this work, thus the number of angular bins ranges between $N_v=\mathcal{O}(10^3$--$10^4)$.

The breaking of symmetry in the EOMs has been recognized and well understood in the context of slow conversion~\cite{Duan:2013kba}, but never before in the presence of MNR. In order to cross-check that the symmetry breaking does not stem from inadequate angular binning, we have varied the angular resolution of the isotropic scenario in Fig.~\ref{fig:6a}. The difference between results in Fig.~\ref{fig:13a} and those in Fig.~\ref{fig:6a} is that in this Appendix we use double precision in the numerical runs instead of the $500$ bits per number implemented before; thus, here we show that the breaking of symmetry is not related to the lack of angular resolution,  rather to the introduction of a small but important floating-point error from the double precision assignment.

In Fig.~\ref{fig:13a}, we show the flavor evolution of $\rho_{ex}$ for the isotropic scenario in the bottom panels of Fig.~\ref{fig:6a}. Notably, the breaking of angular symmetry still happens around $t\simeq 0.3\times10^{-4}$ s. Although not shown, we have repeated the calculation presented in Fig.~\ref{fig:13a} for the solvers Bulirch-Stoer, Rosenbrock, Runge-Kutta-Dopri(5), Runge-Kutta-Cash-Karp(5,4) from the \texttt{Boost} library~\cite{BoostLibrary} as well as using VCABM, DP8, Tsit5 solvers from \texttt{DifferentialEquations.jl} package of \texttt{Julia}~\cite{DifferentialEquations.jl-2017, Julia-2017}, and found that the breaking of symmetry occurs at the same time regardless of the choice of the solver.

For the perturbed and non-perturbed isotropic systems in Sec~\ref{sec:sa}, we implement runs with 500 bits per number and $N_v=100$ angular bins. On the contrary, for the non-isotropic runs of Sec.~\ref{sec:ma-noniso}, since less precision is needed in the floating point error, double precision is enough ({\color{blue}64} bits per number) and one can increase the number of angular bins to $N_{v}=2000$. For the BH disk merger calculations in Sec.~\ref{sec:BH_evolution}, we use double precision and $N_v=30000$ to ensure a reliable final flavor configuration.

\section{Behavior of the neutrino monopole and dipole  due to the matter-neutrino resonance}\label{sec:mono_dipo}

In this appendix, we describe how symmetry-breaking solutions arise and under which conditions they can be expected in different scenarios. In Appendix~\ref{sec:modes}, we investigate the instability conditions of the monopole and dipole modes in the simplest of cases where the matter potential is effectively zero. In Appendix~\ref{sec:lsa_l0l1}, we present the linear stability analysis for the monopole and dipole and compare the analytical predictions of the growth rates to our numerical results.

\subsection{Growth of the monopole and dipole}\label{sec:modes}

We show how the monopole and the dipole modes trigger flavor instabilities in opposite neutrino mass orderings for isotropic systems, building on the work of Ref.~\cite{Duan:2013kba}, we compare the analytical estimates previously known from Ref.~\cite{Duan:2013kba} with our results. 

For simplicity, we introduce the polarization vectors  defined as
\begin{eqnarray}\label{eq:pvectors}
    \vect{P}_v &=& \big(2\mathrm{Re}[\rho_{ex}(v)], -2\mathrm{Im}[\rho_{ex}(v)], \rho_{ee}(v)-\rho_{xx}(v)\big) \ ,
\end{eqnarray}
and similar for $\vect{\bar{P}}_v $ with $\rho_{ij}(v)\rightarrow \bar{\rho}_{ij}(v)$. The EOMs in terms of the sum and  difference of the polarization vectors, $\vect{S}_v=\vect{P}_v+\vect{\bar{P}}_v$ and $\vect{D}_v=\vect{P}_v-\vect{\bar{P}}_v$, are given by
\begin{eqnarray}\label{eq:Dv}
    \dot{\vect{D}}_v &=& -\omega\vect{B}\times\vect{S}_v -\mu\vect{D}_v\times \int (1-v\cdot v^\prime)\vect{D}_{v^\prime}d\Omega^\prime \ , \\
    \dot{\vect{S}}_v &=& -\omega\vect{B}\times\vect{D}_v -\mu\vect{S}_v\times \int (1-v\cdot v^\prime)\vect{D}_{v^\prime}d\Omega^\prime \ ,
    \label{eq:Sv}
\end{eqnarray}
where $d\Omega=dv d\phi$ is the differential solid angle. Although in the main text we have assumed axial symmetry, here we keep the $\phi$ angle to show that both monopole and dipole modes can form regardless of the presence of axial symmetry. 
Moreover, the vacuum potential is represented by the vector $\vect{B}=(\sin{2 \theta_V}, 0, \cos{2\theta_V})$ and we work under the assumption that the matter term is so large that $\theta_V \ll 1$~\cite{Duan:2005cp, Hannestad:2006nj} so that $\vect{B}\approx \vect{e}_z$.

In the linear regime,  Eqs.~\ref{eq:Dv} and~\ref{eq:Sv} can be approximated as
\begin{eqnarray}\label{eq:Dvapprox}
    \dot{\vect{D}}_v &\approx& -\omega\vect{B}\times\vect{S}_v \ , \\
    \dot{\vect{S}}_v &\approx& \Big[ \omega\vect{D}_v + \mu \int (1-v\cdot v^\prime)\vect{D}_{v^\prime}d\Omega^\prime \Big] \times \vect{B} \ .
    \label{eq:Svapprox}
\end{eqnarray}
Here, we have dropped the second term in Eq.~\ref{eq:Dv} because of negligible $\mathcal{O}(|\vect{D}_v|^2)$ and used  $\vect{S}_v \approx \vect{B} \approx \vect{e}_z$. Notice that Eqs.~\ref{eq:Dvapprox} and~\ref{eq:Svapprox} are equivalent to Eqs.~(6a) and (6b) of Ref.~\cite{Duan:2013kba} under the transformation $\vect{S}_v \rightarrow \vect{S}_v - \vect{e}_z \equiv \vect{q}_v$, which leaves the evolution equations unchanged.

To unveil the spontaneous breaking of angular symmetry is useful to express the polarization vectors in terms of the spherical harmonics $Y_{\ell m} (v)$:
\begin{eqnarray}\label{eq:Dlm}
    \tilde{\vect{D}}_{\ell m} &=& \int Y_{\ell m}^{*}(v)\vect{D}_v d\Omega \ , \\
    \tilde{\vect{S}}_{\ell m} &=& \int Y_{\ell m}^{*}(v)\vect{S}_v d\Omega \ ,
    \label{eq:Slm}
\end{eqnarray}
where  $\ell=0,1,...$ and $m=-\ell,...,\ell$. To show the decoupling of the different $\ell, m$ modes in the linear regime, we express the factor $(1-v\cdot v^\prime)$ in the basis of spherical harmonics. Hence Eq.~\ref{eq:Svapprox} becomes
\begin{eqnarray}
    \dot{\vect{S}}_v &\approx& \Big[ \omega\vect{D}_v + 4\pi\mu \int 
\Big( Y_{00}(v)Y_{00}^{*}(v^\prime)-\frac{1}{3}\sum_{m=0,\pm 1} Y_{1m}^{*}(v) Y_{1m}^{*}(v^\prime) \Big) \vect{D}_{v^\prime}d\Omega^\prime \Big] \times \vect{B} \ ,
\end{eqnarray}
involving  the $\ell=0$ and $\ell=1$ spherical harmonics only. To turn this into an equation for the $\ell, m$ modes we multiply both sides by $Y_{\ell m}^{*}(v)$, integrate over $d\Omega$, and use the orthonormality of the spherical harmonics,$\int d\Omega Y_{\ell m}^{*} Y_{\ell^\prime m^\prime} = \delta_{\ell \ell^\prime} \delta_{m m^\prime}$, to obtain: 
\begin{figure*}
\includegraphics[width=0.49\textwidth]{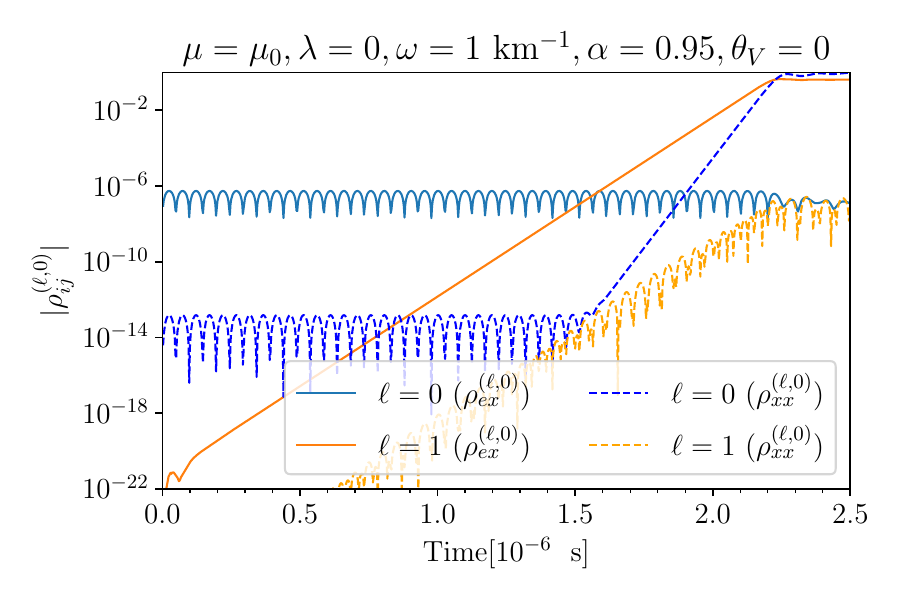}
\includegraphics[width=0.49\textwidth]{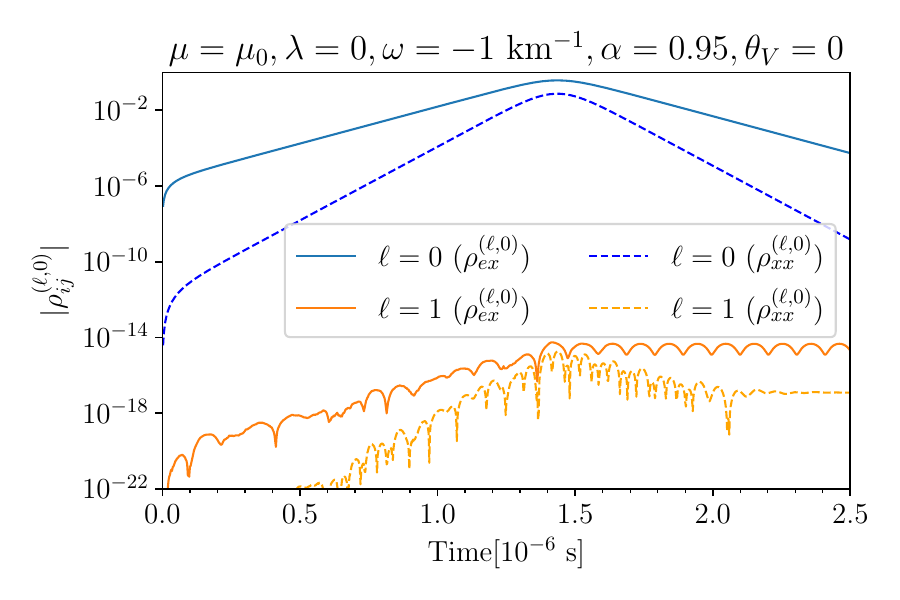}
\caption{Flavor evolution in the normal mass ordering ($\omega=1 
 \ \mathrm{km^{-1}}$, left panel) and inverted mass ordering ($\omega=-1 \ \mathrm{km^{-1}}$, right panel) for a constant $\mu_0=3\times10^3 \ \mathrm{km^{-1}}$. The matter potential is set to $\lambda=0$, we also assume a vanishing mixing angle $\theta_V$ and seed the system with a small off-diagonal seed  [$\mathcal{O}(10^{-8})$]. This choice of parameters is a simplified version of the parameters used in Sec.~\ref{sec:mnr} where $\mu$ does not change in time. In the case with $\omega=1 \ \mathrm{km^{-1}}$, the dipole mode $\ell=1$ (solid orange) exponentially develops with a well-defined growth rate, while the monopole $\ell=0$ (solid blue) is stable. In both left and right panels, the growth rate of the monopole of $\rho_{xx}$ (blue dashed) is exactly twice the growth rate of the unstable mode (in the case with $\omega=-1 \  \mathrm{km^{-1}}$, the unstable mode is the monopole). This is because in the linear regime  $\rho_{xx}\sim \rho_{ex}^2$ (see Sec.~2.2 of Ref.~\cite{Tamborra:2020cul}). 
}
\label{fig:8}
\end{figure*}
\begin{eqnarray}
    \dot{\tilde{\vect{S}}}_{\ell m} \approx \Big[ \omega\tilde{\vect{D}}_{\ell m} + 4\pi\mu \Big(\delta_{\ell 0} \tilde{\vect{D}}_{0 m} - \frac{\delta_{\ell 1} }{3}\tilde{\vect{D}}_{1 m} \Big) \Big] \times \vect{B} \ ,
\end{eqnarray}
or more compactly
\begin{eqnarray}
    \dot{\tilde{\vect{S}}}_{\ell m} \approx \Big[ \omega + \tilde{\mu}_\ell \Big] \tilde{\vect{D}}_{l m}\times \vect{B} \ ,
\end{eqnarray}
where we have defined the $\ell-$dependent interaction strength $\tilde{\mu}_\ell$ as
\begin{equation*}\label{eq:muell}
\tilde{\mu}_\ell = 4\pi \begin{cases}
 \mu &\text{for $\ell=0$} \ , \\
 -\mu/3 &\text{for $\ell=1$} \ , \\
 0 &\text{for $\ell>1$} \ .
\end{cases}
\end{equation*}
The expression above implies that  $\ell=0$ and $\ell=1$ \textit{independently} follow the same EOMs with their corresponding neutrino-neutrino interaction strength given by $\mu$ and $-\mu/3$, respectively~\cite{Duan:2013kba}. Notice that the linearized equations of motion cannot capture the growth of $\ell>1$ modes because they start growing much later when the system is already in the non-linear regime.

The equivalent equation for $\dot{\tilde{\vect{D}}}_{\ell m}$ follows from repeating the same procedure for Eq.~\ref{eq:Dvapprox}. Therefore, the complete set of equations is then
\begin{eqnarray}\label{eq:SDlm}
    \dot{\tilde{\vect{D}}}_{\ell m} &\approx& \omega \tilde{\vect{S}}_{\ell m} \times \vect{B} \ , \nonumber \\
    \dot{\tilde{\vect{S}}}_{\ell m} &\approx& \Big[ \omega + \tilde{\mu}_\ell \Big] \tilde{\vect{D}}_{l m}\times \vect{B} \ .
\end{eqnarray}
In Fig.~\ref{fig:8}, we show how the modes $\ell=0,1$ are decoupled from each other due to the $\ell-$dependence of $\tilde{\mu}_{\ell}$ in the linear regime, which results in a re-scaling and a flip of sign in the neutrino-neutrino interaction strength (see Eq.~\ref{eq:muell}). In Appendix~\ref{sec:lsa_l0l1}, we compute the analytical growth rates of the monopole and the dipole modes and show the agreement with our numerical simulations of flavor evolution.

\subsection{Linear stability analysis for the $\ell=0$ and $\ell=1$ modes}\label{sec:lsa_l0l1}
We carry out the linear stability analysis for the monopole and dipole to investigate the symmetry breaking.\\

{\bf $\ell=0$ mode: The monopole}\\ 

The equations for the monopole $\ell=0$ are given by (see Eqs.~\ref{eq:SDlm}):
\begin{eqnarray}\label{eq:vDdot}
    \dot{\vD} &=& \omega \vS \times \vect{B} \ ,  \\
    \dot{\vS} &=& \omega \vD \times \vect{B} + \mu \vD \times \vS \ ,
    \label{eq:vSdot}
\end{eqnarray}
where $\int dv \vS_v = \vS$ and $ \int dv \vD_v = \vD$ are the angle-integrated vectors of the sum and difference total polarization vector, respectively. We seek to find the eigenfrequency of the system described by the equations above. The transverse components of $\dot{\vD}$ obey the following equations:
\begin{eqnarray}\label{eq:Dxy}
    \dot{D}^x &=& \omega S^y \ , \nonumber  \\
    \dot{D}^y &=&  -\omega S^x \ ,
\end{eqnarray}
and similarly, the ones for $\dot{\vS}$ are given by
\begin{eqnarray}\label{eq:Sxy}
    \dot{S}^x &=&  \mu(-D^z S^y + D^y S^z) + \omega D^y, \nonumber \\
    \dot{S}^y &=&  \mu(D^z S^x - D^x S^z) - \omega D^x \ .
\end{eqnarray}
Using the following linear combinations
\begin{eqnarray}
    \epsilon_S &=& S^{x} - i S^{y} \ , \nonumber  \\
    \epsilon_D &=& D^{x} - i D^{y} \ ,
\end{eqnarray}

\begin{figure*}
\includegraphics[width=0.49\textwidth]{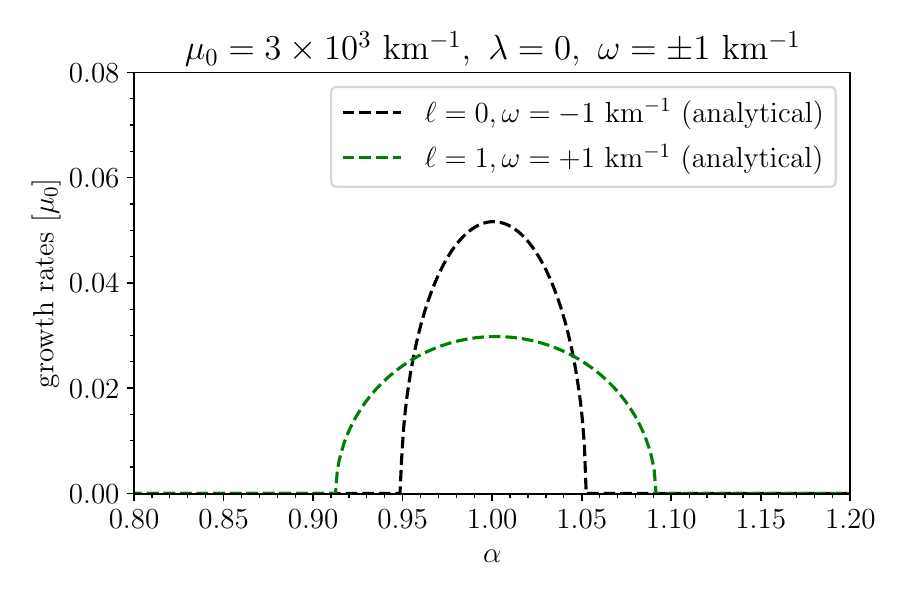}
\includegraphics[width=0.49\textwidth]{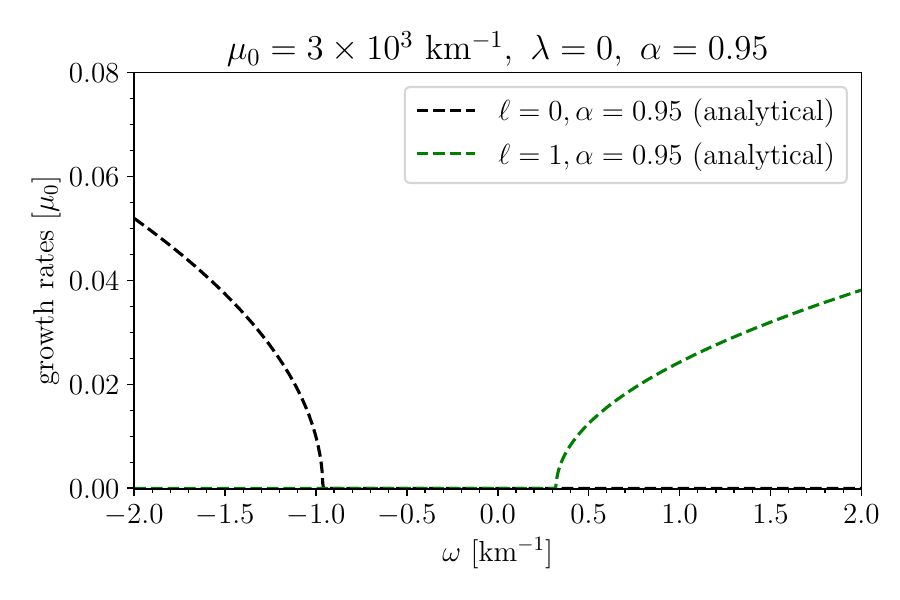}
\includegraphics[width=0.49\textwidth]{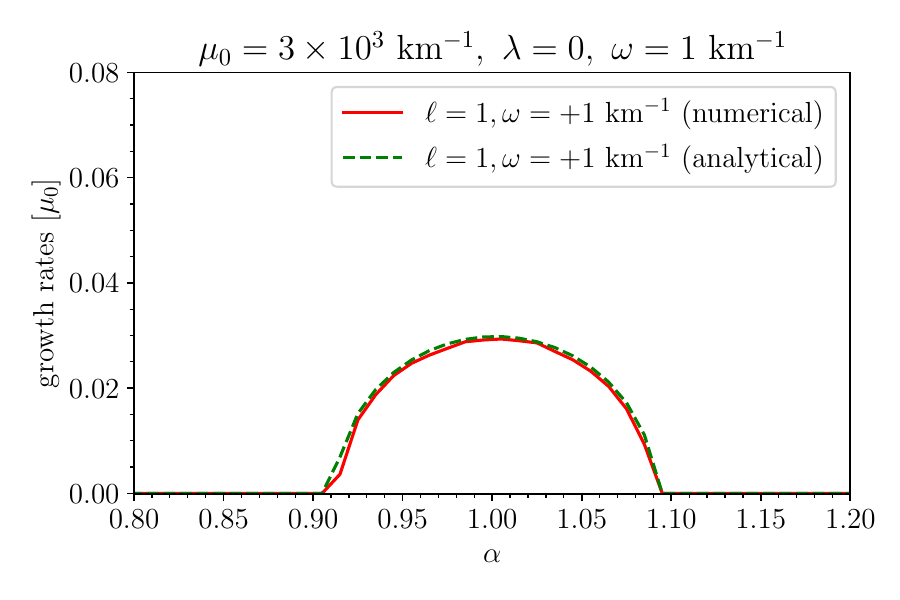}
\includegraphics[width=0.49\textwidth]{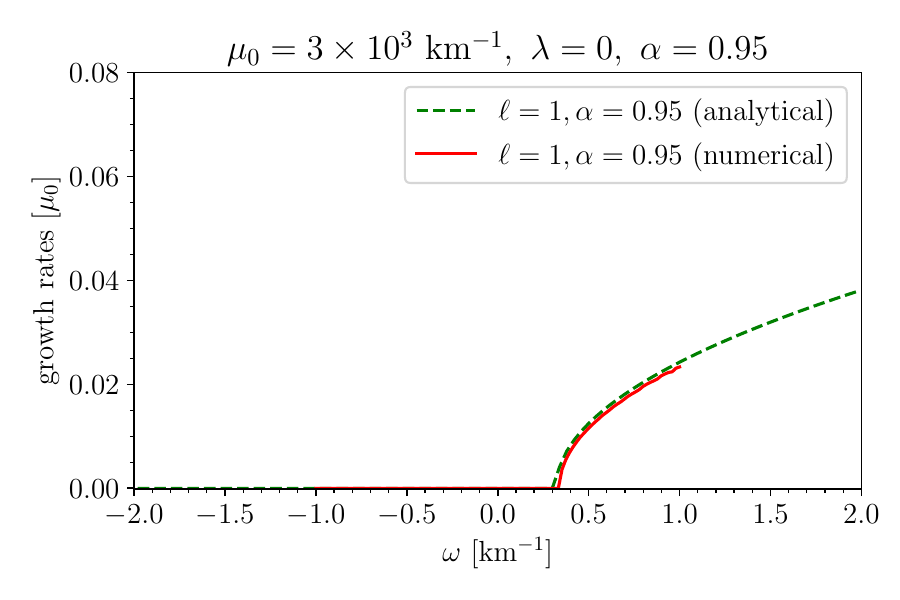}
\caption{
\textit{Top:} Growth rates for the  monopole ($\ell=0$) and dipole ($\ell=1$)  as analytically computed  (Eqs.~\ref{eq:Omegal0} and~\ref{eq:Omegal1}). \textit{Bottom: } Comparison between the prediction of the dipole growth rate (analytical) and the growth rate computed by solving the EOMs (numerical). All cases assume  $\theta_V=0$. This figure shows that solutions that break isotropy (unstable $\ell=1$) are possible only for $\omega>0$ provided that $\mu>0$. As discussed in Sec.~\ref{sec:mnr},  the MNR provides the conditions for isotropic-breaking solutions to occur for both mass orderings, i.e.~the green dashed lines (left) will extend to negative $\omega$.}
\label{fig:9}
\end{figure*}

and combining Eqs.~\ref{eq:Dxy} and~\ref{eq:Sxy} we can obtain a system of equations for the transverse components $\epsilon_S$ and $\epsilon_D$. The linearization assumes that the neutrino transverse component $\epsilon_{D,S}$ is much smaller than $S^{z}$ and $D^z$ so that higher order terms such as  $\epsilon_{S,D}^2$  can be effectively neglected. Thus, the equations of motion for the transverse components are
\begin{eqnarray}
   \dot{\epsilon}_S  &=& -i \mu D^z \epsilon_S + i (\mu S^z + \omega)\epsilon_D \ , \nonumber \\
   \dot{\epsilon}_D  &=& i \omega \epsilon_S \ .
\end{eqnarray}
This set of equations can be expressed in matrix form as
\begin{eqnarray}\label{eq:epsdot}
\begin{bmatrix}
\dot{\epsilon}_S \\
\dot{\epsilon}_D 
\end{bmatrix}
=
(-i) M
\begin{bmatrix}
\epsilon_S \\
\epsilon_D
\end{bmatrix} \ ,
\end{eqnarray}
where the matrix $M$ is defined as 
\begin{eqnarray}
M = 
\begin{bmatrix}
\mu D^z & -\mu S^z -\omega  \\
-\omega & 0 
\end{bmatrix}\ .
\end{eqnarray}
The eigenvalues of $M$ are:
\begin{eqnarray}\label{eq:Omegal0}
    \Omega^{\pm}_{\ell=0} = \frac{1}{2}\mu D^z \pm \frac{1}{2} \sqrt{ (D^z)^2\mu^2 + 4 S^z\mu\omega + 4\omega^2 } \ .
\end{eqnarray}
According to Eq.~\ref{eq:epsdot}, an imaginary $\Omega^\pm_{\ell=0}$ would imply that $\epsilon_{S,D}\propto \exp(\pm\Omega t)$ indicating the existence of an unstable solution in flavor space. The vacuum oscillation frequency $\omega$ must be negative for $\Omega^{\pm}_{\ell=0}$ to have an imaginary component, otherwise the argument of the square root is always positive provided that $\mu>0$. We show the analytical behavior of the root $\Omega^{\pm}_{\ell=0}$ in Fig.~\ref{fig:9} (top panels), where one can see that the root displays an unstable solution only in the inverted mass hierarchy for positive $\mu$. \\

{\bf $\ell=1$ mode: The dipole}\\

We can use the growth rate of the monopole  (Appendix~\ref{sec:modes}) to obtain the growth rate of the dipole by re-defining the neutrino-neutrino interaction strength. The growth rate of the dipole is given by Eq.~\ref{eq:Omegal0} under the transformation $\mu \rightarrow -\mu/3$, namely
\begin{eqnarray}\label{eq:Omegal1}
    \Omega^{\pm}_{\ell=1} = -\frac{1}{6}\mu D^z \pm \frac{1}{2} \sqrt{\frac{1}{9}(D^z)^2\mu^2 - \frac{4}{3} S^z\mu\omega + 4\omega^2 }  \ .
\end{eqnarray}
Note that the transformation $\mu \rightarrow -\mu/3 $ is equivalent to simultaneously changing $\mu \rightarrow \mu/3 $ and $\omega \rightarrow -\omega $. For this reason, the growth rates of the $\ell=1$ mode are smaller than those of $\ell=0$ (re-scaled interaction strength) and occur in the $\omega>0$ range, as opposed to the monopole mode. 

In Fig.~\ref{fig:9} we show show the growth rates $\Omega^{+}_{\ell=0}$ and $\Omega^{+}_{\ell=1}$. Notably, when solutions are stable in the $\ell=0$ mode, the system could still develop a growing $\ell=1$ mode, therefore spontaneously breaking the symmetry of the initial angular distributions~\cite{Raffelt:2013isa, Raffelt:2013rqa, Duan:2013kba, Shalgar:2021oko}. Moreover, in the bottom panels of Fig.~\ref{fig:9} we compare the analytical growth rates $\Omega^{\pm}_{\ell=1}$ to the growth rates of our flavor evolution which have been extracted using the finite difference method; the agreement between our simulations and the analytical estimates is excellent.

\end{appendix}


\bibliographystyle{JHEP}
\bibliography{references.bib}

\providecommand{\href}[2]{#2}\begingroup\raggedright\begin{thebibliography}{10}

\bibitem{Burns:2019byj}
E.~Burns, \emph{{Neutron Star Mergers and How to Study Them}},
  \href{https://doi.org/10.1007/s41114-020-00028-7}{\emph{Living Rev. Rel.}
  {\bfseries 23} (2020) 4}, [\href{https://arxiv.org/abs/1909.06085}{{\ttfamily
  1909.06085}}].

\bibitem{Janka:2022krt}
H.-T. Janka and A.~Bauswein, \emph{{Dynamics and Equation of State Dependencies
  of Relevance for Nucleosynthesis in Supernovae and Neutron Star Mergers}},
  pp.~1--98.
\newblock Springer Nature Singapore, Singapore.
\newblock \href{https://arxiv.org/abs/2212.07498}{{\ttfamily 2212.07498}}.

\bibitem{Metzger:2019zeh}
B.~D. Metzger, \emph{{Kilonovae}},
  \href{https://doi.org/10.1007/s41114-019-0024-0}{\emph{Living Rev. Rel.}
  {\bfseries 23} (2020) 1}, [\href{https://arxiv.org/abs/1910.01617}{{\ttfamily
  1910.01617}}].

\bibitem{Wolfenstein:1977ue}
L.~Wolfenstein, \emph{{Neutrino Oscillations in Matter}},
  \href{https://doi.org/10.1103/PhysRevD.17.2369}{\emph{Phys. Rev.} {\bfseries
  D17} (1978) 2369--2374}.

\bibitem{Mikheev:1986gs}
S.~P. Mikheyev and A.~{\relax Yu}. Smirnov, \emph{{Resonance Amplification of
  Oscillations in Matter and Spectroscopy of Solar Neutrinos}}, {\emph{Sov. J.
  Nucl. Phys.} {\bfseries 42} (1985) 913--917}.

\bibitem{Duan:2010bg}
H.~Duan, G.~M. Fuller and Y.-Z. Qian, \emph{{Collective Neutrino
  Oscillations}},
  \href{https://doi.org/10.1146/annurev.nucl.012809.104524}{\emph{Ann. Rev.
  Nucl. Part. Sci.} {\bfseries 60} (2010) 569--594},
  [\href{https://arxiv.org/abs/1001.2799}{{\ttfamily 1001.2799}}].

\bibitem{Mirizzi:2015eza}
A.~Mirizzi, I.~Tamborra, H.-T. Janka, N.~Saviano, K.~Scholberg, R.~Bollig
  et~al., \emph{{Supernova Neutrinos: Production, Oscillations and Detection}},
  \href{https://doi.org/10.1393/ncr/i2016-10120-8}{\emph{Riv. Nuovo Cim.}
  {\bfseries 39} (2016) 1--112},
  [\href{https://arxiv.org/abs/1508.00785}{{\ttfamily 1508.00785}}].

\bibitem{Tamborra:2020cul}
I.~Tamborra and S.~Shalgar, \emph{{New Developments in Flavor Evolution of a
  Dense Neutrino Gas}},
  \href{https://doi.org/10.1146/annurev-nucl-102920-050505}{\emph{Ann. Rev.
  Nucl. Part. Sci.} {\bfseries 71} (2021) 165--188},
  [\href{https://arxiv.org/abs/2011.01948}{{\ttfamily 2011.01948}}].

\bibitem{Richers:2022zug}
S.~Richers and M.~Sen, \emph{{Fast Flavor Transformations}}, pp.~1--17.
\newblock Springer Nature Singapore, Singapore.
\newblock \href{https://arxiv.org/abs/2207.03561}{{\ttfamily 2207.03561}}.

\bibitem{Chakraborty:2016yeg}
S.~Chakraborty, R.~Hansen, I.~Izaguirre and G.~G. Raffelt, \emph{{Collective
  neutrino flavor conversion: Recent developments}},
  \href{https://doi.org/10.1016/j.nuclphysb.2016.02.012}{\emph{Nucl. Phys.}
  {\bfseries B908} (2016) 366--381},
  [\href{https://arxiv.org/abs/1602.02766}{{\ttfamily 1602.02766}}].

\bibitem{Malkus:2012ts}
A.~Malkus, J.~P. Kneller, G.~C. McLaughlin and R.~Surman, \emph{{Neutrino
  oscillations above black hole accretion disks: disks with electron-flavor
  emission}}, \href{https://doi.org/10.1103/PhysRevD.86.085015}{\emph{Phys.
  Rev. D} {\bfseries 86} (2012) 085015},
  [\href{https://arxiv.org/abs/1207.6648}{{\ttfamily 1207.6648}}].

\bibitem{Malkus:2014iqa}
A.~Malkus, A.~Friedland and G.~C. McLaughlin, \emph{{Matter-Neutrino Resonance
  Above Merging Compact Objects}},
  \href{https://arxiv.org/abs/1403.5797}{{\ttfamily 1403.5797}}.

\bibitem{Malkus:2015mda}
A.~Malkus, G.~C. McLaughlin and R.~Surman, \emph{{Symmetric and Standard
  Matter-Neutrino Resonances Above Merging Compact Objects}},
  \href{https://doi.org/10.1103/PhysRevD.93.045021}{\emph{Phys. Rev. D}
  {\bfseries 93} (2016) 045021},
  [\href{https://arxiv.org/abs/1507.00946}{{\ttfamily 1507.00946}}].

\bibitem{Wu:2015fga}
M.-R. Wu, H.~Duan and Y.-Z. Qian, \emph{{Physics of neutrino flavor
  transformation through matter--neutrino resonances}},
  \href{https://doi.org/10.1016/j.physletb.2015.11.027}{\emph{Phys. Lett. B}
  {\bfseries 752} (2016) 89--94},
  [\href{https://arxiv.org/abs/1509.08975}{{\ttfamily 1509.08975}}].

\bibitem{Zhu:2016mwa}
Y.-L. Zhu, A.~Perego and G.~C. McLaughlin, \emph{{Matter Neutrino Resonance
  Transitions above a Neutron Star Merger Remnant}},
  \href{https://doi.org/10.1103/PhysRevD.94.105006}{\emph{Phys. Rev. D}
  {\bfseries 94} (2016) 105006},
  [\href{https://arxiv.org/abs/1607.04671}{{\ttfamily 1607.04671}}].

\bibitem{Vaananen:2015hfa}
D.~V{\"a}{\"a}n{\"a}nen and G.~C. McLaughlin, \emph{{Uncovering the
  Matter-Neutrino Resonance}},
  \href{https://doi.org/10.1103/PhysRevD.93.105044}{\emph{Phys. Rev. D}
  {\bfseries 93} (2016) 105044},
  [\href{https://arxiv.org/abs/1510.00751}{{\ttfamily 1510.00751}}].

\bibitem{Frensel:2016fge}
M.~Frensel, M.-R. Wu, C.~Volpe and A.~Perego, \emph{{Neutrino Flavor Evolution
  in Binary Neutron Star Merger Remnants}},
  \href{https://doi.org/10.1103/PhysRevD.95.023011}{\emph{Phys. Rev. D}
  {\bfseries 95} (2017) 023011},
  [\href{https://arxiv.org/abs/1607.05938}{{\ttfamily 1607.05938}}].

\bibitem{Tian:2017xbr}
J.~Y. Tian, A.~V. Patwardhan and G.~M. Fuller, \emph{{Neutrino Flavor Evolution
  in Neutron Star Mergers}},
  \href{https://doi.org/10.1103/PhysRevD.96.043001}{\emph{Phys. Rev. D}
  {\bfseries 96} (2017) 043001},
  [\href{https://arxiv.org/abs/1703.03039}{{\ttfamily 1703.03039}}].

\bibitem{Shalgar:2017pzd}
S.~Shalgar, \emph{{Multi-angle calculation of the matter-neutrino resonance
  near an accretion disk}},
  \href{https://doi.org/10.1088/1475-7516/2018/02/010}{\emph{JCAP} {\bfseries
  02} (2018) 010}, [\href{https://arxiv.org/abs/1707.07692}{{\ttfamily
  1707.07692}}].

\bibitem{Vlasenko:2018irq}
A.~Vlasenko and G.~C. McLaughlin, \emph{{Matter-neutrino resonance in a
  multiangle neutrino bulb model}},
  \href{https://doi.org/10.1103/PhysRevD.97.083011}{\emph{Phys. Rev. D}
  {\bfseries 97} (2018) 083011},
  [\href{https://arxiv.org/abs/1801.07813}{{\ttfamily 1801.07813}}].

\bibitem{Sigurdarson:2022mcm}
G.~{Sigurdharson}, I.~Tamborra and M.-R. Wu, \emph{{Resonant production of
  light sterile neutrinos in compact binary merger remnants}},
  \href{https://doi.org/10.1103/PhysRevD.106.123030}{\emph{Phys. Rev. D}
  {\bfseries 106} (2022) 123030},
  [\href{https://arxiv.org/abs/2209.07544}{{\ttfamily 2209.07544}}].

\bibitem{Cirigliano:2017hmk}
V.~Cirigliano, M.~W. Paris and S.~Shalgar, \emph{{Effect of collisions on
  neutrino flavor inhomogeneity in a dense neutrino gas}},
  \href{https://doi.org/10.1016/j.physletb.2017.09.039}{\emph{Phys. Lett. B}
  {\bfseries 774} (2017) 258--267},
  [\href{https://arxiv.org/abs/1706.07052}{{\ttfamily 1706.07052}}].

\bibitem{Just:2014fka}
O.~Just, A.~Bauswein, R.~A. Pulpillo, S.~Goriely and H.-T. Janka,
  \emph{{Comprehensive nucleosynthesis analysis for ejecta of compact binary
  mergers}}, \href{https://doi.org/10.1093/mnras/stv009}{\emph{Mon. Not. Roy.
  Astron. Soc.} {\bfseries 448} (2015) 541--567},
  [\href{https://arxiv.org/abs/1406.2687}{{\ttfamily 1406.2687}}].

\bibitem{Just:2015dba}
O.~Just, M.~Obergaulinger, H.-T. Janka, A.~Bauswein and N.~Schwarz,
  \emph{{Neutron-star merger ejecta as obstacles to neutrino-powered jets of
  gamma-ray bursts}},
  \href{https://doi.org/10.3847/2041-8205/816/2/L30}{\emph{Astrophys. J. Lett.}
  {\bfseries 816} (2016) L30},
  [\href{https://arxiv.org/abs/1510.04288}{{\ttfamily 1510.04288}}].

\bibitem{Janka:1999qu}
H.-T. Janka, T.~Eberl, M.~Ruffert and C.~L. Fryer, \emph{{Black hole: Neutron
  star mergers as central engines of gamma-ray bursts}},
  \href{https://doi.org/10.1086/312397}{\emph{Astrophys. J. Lett.} {\bfseries
  527} (1999) L39}, [\href{https://arxiv.org/abs/astro-ph/9908290}{{\ttfamily
  astro-ph/9908290}}].

\bibitem{Sekiguchi:2016bjd}
Y.~Sekiguchi, K.~Kiuchi, K.~Kyutoku, M.~Shibata and K.~Taniguchi,
  \emph{{Dynamical mass ejection from the merger of asymmetric binary neutron
  stars: Radiation-hydrodynamics study in general relativity}},
  \href{https://doi.org/10.1103/PhysRevD.93.124046}{\emph{Phys. Rev. D}
  {\bfseries 93} (2016) 124046},
  [\href{https://arxiv.org/abs/1603.01918}{{\ttfamily 1603.01918}}].

\bibitem{Shalgar:2019qwg}
S.~Shalgar, I.~Padilla-Gay and I.~Tamborra, \emph{{Neutrino propagation hinders
  fast pairwise flavor conversions}},
  \href{https://doi.org/10.1088/1475-7516/2020/06/048}{\emph{JCAP} {\bfseries
  06} (2020) 048}, [\href{https://arxiv.org/abs/1911.09110}{{\ttfamily
  1911.09110}}].

\bibitem{Padilla-Gay:2020uxa}
I.~Padilla-Gay, S.~Shalgar and I.~Tamborra, \emph{{Multi-Dimensional Solution
  of Fast Neutrino Conversions in Binary Neutron Star Merger Remnants}},
  \href{https://doi.org/10.1088/1475-7516/2021/01/017}{\emph{JCAP} {\bfseries
  01} (2021) 017}, [\href{https://arxiv.org/abs/2009.01843}{{\ttfamily
  2009.01843}}].

\bibitem{Shalgar:2022rjj}
S.~Shalgar and I.~Tamborra, \emph{{Neutrino decoupling is altered by flavor
  conversion}}, \href{https://doi.org/10.1103/PhysRevD.108.043006}{\emph{Phys.
  Rev. D} {\bfseries 108} (2023) 043006},
  [\href{https://arxiv.org/abs/2206.00676}{{\ttfamily 2206.00676}}].

\bibitem{Shalgar:2022lvv}
S.~Shalgar and I.~Tamborra, \emph{{Neutrino flavor conversion, advection, and
  collisions: Toward the full solution}},
  \href{https://doi.org/10.1103/PhysRevD.107.063025}{\emph{Phys. Rev. D}
  {\bfseries 107} (2023) 063025},
  [\href{https://arxiv.org/abs/2207.04058}{{\ttfamily 2207.04058}}].

\bibitem{Padilla-Gay:2022wck}
I.~Padilla-Gay, I.~Tamborra and G.~G. Raffelt, \emph{{Neutrino fast flavor
  pendulum. II. Collisional damping}},
  \href{https://doi.org/10.1103/PhysRevD.106.103031}{\emph{Phys. Rev. D}
  {\bfseries 106} (2022) 103031},
  [\href{https://arxiv.org/abs/2209.11235}{{\ttfamily 2209.11235}}].

\bibitem{Fiorillo:2023ajs}
D.~F.~G. Fiorillo, I.~Padilla-Gay and G.~G. Raffelt, \emph{{Collisions and
  collective flavor conversion: Integrating out the fast dynamics}},
  \href{https://doi.org/10.1103/PhysRevD.109.063021}{\emph{Phys. Rev. D}
  {\bfseries 109} (2024) 063021},
  [\href{https://arxiv.org/abs/2312.07612}{{\ttfamily 2312.07612}}].

\bibitem{Fogli:2007bk}
G.~L. Fogli, E.~Lisi, A.~Marrone and A.~Mirizzi, \emph{{Collective neutrino
  flavor transitions in supernovae and the role of trajectory averaging}},
  \href{https://doi.org/10.1088/1475-7516/2007/12/010}{\emph{JCAP} {\bfseries
  12} (2007) 010}, [\href{https://arxiv.org/abs/0707.1998}{{\ttfamily
  0707.1998}}].

\bibitem{Duan:2006an}
H.~Duan, G.~M. Fuller, J.~Carlson and Y.-Z. Qian, \emph{{Simulation of Coherent
  Non-Linear Neutrino Flavor Transformation in the Supernova Environment. 1.
  Correlated Neutrino Trajectories}},
  \href{https://doi.org/10.1103/PhysRevD.74.105014}{\emph{Phys. Rev. D}
  {\bfseries 74} (2006) 105014},
  [\href{https://arxiv.org/abs/astro-ph/0606616}{{\ttfamily
  astro-ph/0606616}}].

\bibitem{Shalgar:2021oko}
S.~Shalgar and I.~Tamborra, \emph{{Symmetry breaking induced by pairwise
  conversion of neutrinos in compact sources}},
  \href{https://doi.org/10.1103/PhysRevD.105.043018}{\emph{Phys. Rev. D}
  {\bfseries 105} (2022) 043018},
  [\href{https://arxiv.org/abs/2106.15622}{{\ttfamily 2106.15622}}].

\bibitem{DifferentialEquations.jl-2017}
C.~Rackauckas and Q.~Nie, \emph{Differentialequations.jl – a performant and
  feature-rich ecosystem for solving differential equations in julia},
  \href{https://doi.org/10.5334/jors.151}{\emph{The Journal of Open Research
  Software} {\bfseries 5} (2017) }.

\bibitem{Julia-2017}
J.~Bezanson, A.~Edelman, S.~Karpinski and V.~B. Shah, \emph{Julia: A fresh
  approach to numerical computing},
  \href{https://doi.org/10.1137/141000671}{\emph{SIAM {R}eview} {\bfseries 59}
  (2017) 65--98}.

\bibitem{Duan:2013kba}
H.~Duan, \emph{{Flavor Oscillation Modes In Dense Neutrino Media}},
  \href{https://doi.org/10.1103/PhysRevD.88.125008}{\emph{Phys. Rev. D}
  {\bfseries 88} (2013) 125008},
  [\href{https://arxiv.org/abs/1309.7377}{{\ttfamily 1309.7377}}].

\bibitem{Just_2015}
O.~Just, A.~Bauswein, R.~A. Pulpillo, S.~Goriely and H.-T. Janka,
  \emph{Comprehensive nucleosynthesis analysis for ejecta of compact binary
  mergers}, \href{https://doi.org/10.1093/mnras/stv009}{\emph{Monthly Notices
  of the Royal Astronomical Society} {\bfseries 448} (feb, 2015) 541--567}.

\bibitem{Wu:2017drk}
M.-R. Wu, I.~Tamborra, O.~Just and H.-T. Janka, \emph{{Imprints of
  neutrino-pair flavor conversions on nucleosynthesis in ejecta from
  neutron-star merger remnants}},
  \href{https://doi.org/10.1103/PhysRevD.96.123015}{\emph{Phys. Rev. D}
  {\bfseries 96} (2017) 123015},
  [\href{https://arxiv.org/abs/1711.00477}{{\ttfamily 1711.00477}}].

\bibitem{Murchikova:2017zsy}
L.~M. Murchikova, E.~Abdikamalov and T.~Urbatsch, \emph{{Analytic Closures for
  M1 Neutrino Transport}},
  \href{https://doi.org/10.1093/mnras/stx986}{\emph{Mon. Not. Roy. Astron.
  Soc.} {\bfseries 469} (2017) 1725--1737},
  [\href{https://arxiv.org/abs/1701.07027}{{\ttfamily 1701.07027}}].

\bibitem{Cernohorsky:1994yg}
J.~{Cernohorsky} and S.~A. {Bludman}, \emph{{Maximum Entropy Distribution and
  Closure for Bose-Einstein and Fermi-Dirac Radiation Transport}},
  \href{https://doi.org/10.1086/174640}{\emph{Astrophys. J.} {\bfseries 433}
  (Sept., 1994) 250}.

\bibitem{Richers:2022dqa}
S.~Richers, \emph{{Evaluating approximate flavor instability metrics in neutron
  star mergers}},
  \href{https://doi.org/10.1103/PhysRevD.106.083005}{\emph{Phys. Rev. D}
  {\bfseries 106} (2022) 083005},
  [\href{https://arxiv.org/abs/2206.08444}{{\ttfamily 2206.08444}}].

\bibitem{Izaguirre:2016gsx}
I.~Izaguirre, G.~G. Raffelt and I.~Tamborra, \emph{{Fast Pairwise Conversion of
  Supernova Neutrinos: A Dispersion-Relation Approach}},
  \href{https://doi.org/10.1103/PhysRevLett.118.021101}{\emph{Phys. Rev. Lett.}
  {\bfseries 118} (2017) 021101},
  [\href{https://arxiv.org/abs/1610.01612}{{\ttfamily 1610.01612}}].

\bibitem{Morinaga:2021vmc}
T.~Morinaga, \emph{{Fast neutrino flavor instability and neutrino flavor lepton
  number crossings}},
  \href{https://doi.org/10.1103/PhysRevD.105.L101301}{\emph{Phys. Rev. D}
  {\bfseries 105} (2022) L101301},
  [\href{https://arxiv.org/abs/2103.15267}{{\ttfamily 2103.15267}}].

\bibitem{Padilla-Gay:2021haz}
I.~Padilla-Gay, I.~Tamborra and G.~G. Raffelt, \emph{{Neutrino Flavor Pendulum
  Reloaded: The Case of Fast Pairwise Conversion}},
  \href{https://doi.org/10.1103/PhysRevLett.128.121102}{\emph{Phys. Rev. Lett.}
  {\bfseries 128} (2022) 121102},
  [\href{https://arxiv.org/abs/2109.14627}{{\ttfamily 2109.14627}}].

\bibitem{Wu:2017qpc}
M.-R. Wu and I.~Tamborra, \emph{{Fast neutrino conversions: Ubiquitous in
  compact binary merger remnants}},
  \href{https://doi.org/10.1103/PhysRevD.95.103007}{\emph{Phys. Rev. D}
  {\bfseries 95} (2017) 103007},
  [\href{https://arxiv.org/abs/1701.06580}{{\ttfamily 1701.06580}}].

\bibitem{Just:2022flt}
O.~Just, S.~Abbar, M.-R. Wu, I.~Tamborra, H.-T. Janka and F.~Capozzi,
  \emph{{Fast neutrino conversion in hydrodynamic simulations of
  neutrino-cooled accretion disks}},
  \href{https://doi.org/10.1103/PhysRevD.105.083024}{\emph{Phys. Rev. D}
  {\bfseries 105} (2022) 083024},
  [\href{https://arxiv.org/abs/2203.16559}{{\ttfamily 2203.16559}}].

\bibitem{George:2020veu}
M.~George, M.-R. Wu, I.~Tamborra, R.~Ardevol-Pulpillo and H.-T. Janka,
  \emph{{Fast neutrino flavor conversion, ejecta properties, and
  nucleosynthesis in newly-formed hypermassive remnants of neutron-star
  mergers}}, \href{https://doi.org/10.1103/PhysRevD.102.103015}{\emph{Phys.
  Rev. D} {\bfseries 102} (2020) 103015},
  [\href{https://arxiv.org/abs/2009.04046}{{\ttfamily 2009.04046}}].

\bibitem{Li:2021vqj}
X.~Li and D.~M. Siegel, \emph{{Neutrino Fast Flavor Conversions in Neutron-Star
  Postmerger Accretion Disks}},
  \href{https://doi.org/10.1103/PhysRevLett.126.251101}{\emph{Phys. Rev. Lett.}
  {\bfseries 126} (2021) 251101},
  [\href{https://arxiv.org/abs/2103.02616}{{\ttfamily 2103.02616}}].

\bibitem{Fernandez:2022yyv}
R.~Fern\'andez, S.~Richers, N.~Mulyk and S.~Fahlman, \emph{{Fast flavor
  instability in hypermassive neutron star disk outflows}},
  \href{https://doi.org/10.1103/PhysRevD.106.103003}{\emph{Phys. Rev. D}
  {\bfseries 106} (2022) 103003},
  [\href{https://arxiv.org/abs/2207.10680}{{\ttfamily 2207.10680}}].

\bibitem{Johns:2021qby}
L.~Johns, \emph{{Collisional Flavor Instabilities of Supernova Neutrinos}},
  \href{https://doi.org/10.1103/PhysRevLett.130.191001}{\emph{Phys. Rev. Lett.}
  {\bfseries 130} (2023) 191001},
  [\href{https://arxiv.org/abs/2104.11369}{{\ttfamily 2104.11369}}].

\bibitem{Johns:2022yqy}
L.~Johns and Z.~Xiong, \emph{{Collisional instabilities of neutrinos and their
  interplay with fast flavor conversion in compact objects}},
  \href{https://doi.org/10.1103/PhysRevD.106.103029}{\emph{Phys. Rev. D}
  {\bfseries 106} (2022) 103029},
  [\href{https://arxiv.org/abs/2208.11059}{{\ttfamily 2208.11059}}].

\bibitem{Xiong:2022zqz}
Z.~Xiong, L.~Johns, M.-R. Wu and H.~Duan, \emph{{Collisional flavor instability
  in dense neutrino gases}},
  \href{https://doi.org/10.1103/PhysRevD.108.083002}{\emph{Phys. Rev. D}
  {\bfseries 108} (2023) 083002},
  [\href{https://arxiv.org/abs/2212.03750}{{\ttfamily 2212.03750}}].

\bibitem{BoostLibrary}
Boost, ``{Boost C++ Libraries}.'' \url{http://www.boost.org/}, 2015.

\bibitem{Sarikas:2012ad}
S.~Sarikas, D.~de~Sousa~Seixas and G.~G. Raffelt, \emph{{Spurious instabilities
  in multi-angle simulations of collective flavor conversion}},
  \href{https://doi.org/10.1103/PhysRevD.86.125020}{\emph{Phys. Rev. D}
  {\bfseries 86} (2012) 125020},
  [\href{https://arxiv.org/abs/1210.4557}{{\ttfamily 1210.4557}}].

\bibitem{Morinaga:2018aug}
T.~Morinaga and S.~Yamada, \emph{{Linear stability analysis of collective
  neutrino oscillations without spurious modes}},
  \href{https://doi.org/10.1103/PhysRevD.97.023024}{\emph{Phys. Rev. D}
  {\bfseries 97} (2018) 023024},
  [\href{https://arxiv.org/abs/1803.05913}{{\ttfamily 1803.05913}}].

\bibitem{Duan:2005cp}
H.~Duan, G.~M. Fuller and Y.-Z. Qian, \emph{{Collective neutrino flavor
  transformation in supernovae}},
  \href{https://doi.org/10.1103/PhysRevD.74.123004}{\emph{Phys. Rev. D}
  {\bfseries 74} (2006) 123004},
  [\href{https://arxiv.org/abs/astro-ph/0511275}{{\ttfamily
  astro-ph/0511275}}].

\bibitem{Hannestad:2006nj}
S.~Hannestad, G.~G. Raffelt, G.~Sigl and Y.~Y.~Y. Wong, \emph{{Self-induced
  conversion in dense neutrino gases: Pendulum in flavour space}},
  \href{https://doi.org/10.1103/PhysRevD.74.105010}{\emph{Phys. Rev. D}
  {\bfseries 74} (2006) 105010},
  [\href{https://arxiv.org/abs/astro-ph/0608695}{{\ttfamily
  astro-ph/0608695}}].

\bibitem{Raffelt:2013isa}
G.~G. Raffelt and D.~de~Sousa~Seixas, \emph{{Neutrino flavor pendulum in both
  mass hierarchies}},
  \href{https://doi.org/10.1103/PhysRevD.88.045031}{\emph{Phys. Rev. D}
  {\bfseries 88} (2013) 045031},
  [\href{https://arxiv.org/abs/1307.7625}{{\ttfamily 1307.7625}}].

\bibitem{Raffelt:2013rqa}
G.~G. Raffelt, S.~Sarikas and D.~de~Sousa~Seixas, \emph{{Axial Symmetry
  Breaking in Self-Induced Flavor Conversion of Supernova Neutrino Fluxes}},
  \href{https://doi.org/10.1103/PhysRevLett.113.239903,
  10.1103/PhysRevLett.111.091101}{\emph{Phys. Rev. Lett.} {\bfseries 111}
  (2013) 091101}, [\href{https://arxiv.org/abs/1305.7140}{{\ttfamily
  1305.7140}}].

\end{thebibliography}\endgroup
\end{document}